\documentclass[aps,pra,twocolumn,amsmath,amssymb,superscriptaddress,longbibliography]{revtex4-1}
\usepackage[english]{babel}

\usepackage{amssymb}
\usepackage{dcolumn}
\usepackage{bm}
\usepackage{graphicx}
\usepackage{amsmath}

\usepackage{graphicx}        
\graphicspath{{pict/}{}}

\usepackage{dcolumn}
\usepackage{bm}
\usepackage[dvipdfm, pdfstartview=FitH, CJKbookmarks=true, bookmarksnumbered=true, bookmarksopen=true, colorlinks=true, pdfborder=001, citecolor=blue, linkcolor=blue, urlcolor=blue, linktocpage=true] {hyperref}

\begin{document}

\title{Topological Floquet edge states in periodically curved waveguides}

\author{Bo Zhu}
\affiliation{Laboratory of Quantum Engineering and Quantum Metrology, School of Physics and Astronomy, Sun Yat-Sen University (Zhuhai Campus), Zhuhai 519082, China}
\affiliation{State Key Laboratory of Optoelectronic Materials and Technologies, Sun Yat-Sen University (Guangzhou Campus), Guangzhou 510275, China}

\author{Honghua Zhong}
\affiliation{Laboratory of Quantum Engineering and Quantum Metrology, School of Physics and Astronomy, Sun Yat-Sen University (Zhuhai Campus), Zhuhai 519082, China}
\affiliation{Institute of Mathematics and Physics, Central South University of Forestry and Technology, Changsha 410004, China}

\author{Yongguan Ke}
\affiliation{Laboratory of Quantum Engineering and Quantum Metrology, School of Physics and Astronomy, Sun Yat-Sen University (Zhuhai Campus), Zhuhai 519082, China}
\affiliation{State Key Laboratory of Optoelectronic Materials and Technologies, Sun Yat-Sen University (Guangzhou Campus), Guangzhou 510275, China}

\author{Xizhou Qin}
\affiliation{Laboratory of Quantum Engineering and Quantum Metrology, School of Physics and Astronomy, Sun Yat-Sen University (Zhuhai Campus), Zhuhai 519082, China}

\author{Andrey A. Sukhorukov}
\author{Yuri S. Kivshar}
\affiliation{Nonlinear Physics Centre, Research School of Physics and Engineering, The Australian National University, Canberra ACT 2601, Australia}

\author{Chaohong Lee}
\altaffiliation{Corresponding author.\\ Email: lichaoh2@mail.sysu.edu.cn, chleecn@gmail.com.}
\affiliation{Laboratory of Quantum Engineering and Quantum Metrology, School of Physics and Astronomy, Sun Yat-Sen University (Zhuhai Campus), Zhuhai 519082, China}
\affiliation{State Key Laboratory of Optoelectronic Materials and Technologies, Sun Yat-Sen University (Guangzhou Campus), Guangzhou 510275, China}
\affiliation{Synergetic Innovation Center for Quantum Effects and Applications, Hunan Normal University, Changsha 410081, China}

\date{\today}

\begin{abstract}
We study the Floquet edge states in arrays of periodically curved optical waveguides described by the modulated Su-Schrieffer-Heeger model.
Beyond the bulk-edge correspondence, our study explores the interplay between band topology and periodic modulations.
By analysing the quasi-energy spectra and Zak phase, we reveal that, although topological and non-topological edge states can exist for the same parameters, \emph{they can not appear in the same spectral gap}.
In the high-frequency limit, we find analytically all boundaries between the different phases and study the coexistence of topological and non-topological edge states.
In contrast to unmodulated systems, the edge states appear due to either band topology or modulation-induced defects.
This means that periodic modulations may not only tune the parametric regions with nontrivial topology, but may also support novel edge states.
\end{abstract}

\maketitle

\section{Introduction\label{Sec1}}
Recently, topological photonics has emerged as a new approach to manipulate properties of light under continuous deformations~\cite{Lu:2014-821:NPHOT}.
Electromagnetic topological states have been found in both microwave~\cite{Haldane:2008-13904:PRL, Raghu:2008-33834:PRA, Wang:2009-772:NAT} and optical~\cite{Umucalilar:2011-43804:PRA, Fang:2012-782:NPHOT, Khanikaev:2013-233:NMAT} regimes.
Similar to topological insulators for electrons, photonic topological insulators have also been created~\cite{Haldane:2008-13904:PRL, Raghu:2008-33834:PRA, Wang:2009-772:NAT, Umucalilar:2011-43804:PRA, Fang:2012-782:NPHOT, Kraus:2012-106402:PRL, Khanikaev:2013-233:NMAT, Rechtsman:2013-196:NAT, Hafezi:2013-1001:NPHOT, Liang:2013-203904:PRL, Lu:2014-821:NPHOT, Pasek:2014-75113:PRB, Hu:2015-11012:PRX, Wu:2015-223901:PRL, He:2016-4924:PNAS, Gao:2016-11619:NCOM}.
Beyond conventional topological phenomena in linear Hermitian systems, topological gap solitons have been found in nonlinear optical systems~\cite{Lumer:2013-243905:PRL}, and it was shown that topological states can survive in non-Hermitian systems~\cite{Zeuner:2015-40402:PRL}.
Moreover, periodic modulations can bring several novel topological properties  usually absent in their non-modulated analogues~\cite{Rechtsman:2013-196:NAT, Rudner:2013-31005:PRX,Gomez-Leon:2013-200403:PRL, Slobozhanyuk:2015-123901:PRL,Poli:2015-6710:NCOM, Leykam:2016-13902:PRL, Ke:2016-995:LPR, Maczewsky:2017-13756:NCOM, Mukherjee:2017-13918:NCOM}.

Bulk-edge correspondence~\cite{Hatsugai:1993-3697:PRL, Hatsugai:1993-11851:PRB} is a well-established principle for two-dimensional (2D) topological systems.
It establishes the exact correspondence between bulk states subjected to periodic boundary conditions (PBCs) and edge states in the systems with open boundary conditions (OBCs).
Up to now, topological edge states have been found in several 2D photonic systems~\cite{Hafezi:2013-1001:NPHOT, Hafezi:2014-210405:PRL, Hu:2015-11012:PRX, Mittal:2016-180:NPHOT}.
However, for one-dimensional (1D) lattice models, edge states have been shown to appear in periodically modulated but non-topological lattices~\cite{Garanovich:2008-203904:PRL, Szameit:2008-203902:PRL}.
This suggests that edge states can be induced by either topology or periodic modulations.
Here, we wonder whether topological and non-topological edge states may coexist and, if they may coexist, how to distinguish between topological and non-topological edge states.

In this work, we study the Floquet edge states (FESs) in arrays of periodically curved optical waveguides described by a periodically modulated Su-Schrieffer-Heeger (SSH) model~\cite{Su:1979-1698:PRL}.
We analyse, for the first time to our knowledge, the interplay between band topology and periodic modulations, and describe the coexistence of both topological and non-topological edge states supported by the same parameters.
Our results show that, for a specific gap, the Zak phase $Z_{G_m}$ is either $0$ or $\pi$, so that the topological edge states appear only in the gap of $Z_{G_m}=\pi$.
Through controlling both modulation frequency and amplitude, we may drive the system from non-topological to topological regime, and vice versa.
We demonstrate analytically that periodic modulations induce \emph{a virtual defect at the boundary}, being the key mechanism for the formation of non-topological edge states.

The paper is organized as follows.
In Sec.~\ref{Sec2}, we introduce our physical model and derive its coupled-mode equations.
In Sec.~\ref{Secss}, we calculate the quasi-energy spectra under OBC.
In Sec.~\ref{Sec3}, by employing the multi-scale perturbation analysis, we give the effective coupled-mode equations and demonstrate the periodic modulations can induce virtual defects at boundaries.
The FESs include defect-free surface states and Shockley-like surface states, which induced by virtual defects and the alternating strong and weak couplings between waveguides, respectively.
In Sec.~\ref{Sec4}, we analytically obtain the asymptotic phase boundary and numerically give the phase digram of appear FESs, respectively.
We explore the topological nature of all FESs via calculate the bulk topological invariant Zak phase.
We find that Shockley-like surface states are topological FESs and defect-free surface states are non-topological FESs.
A brief summary is given in Sec.~\ref{Sec5}.

\section{Model\label{Sec2}}

\begin{figure}[htp]
\center
\includegraphics[width=\columnwidth]{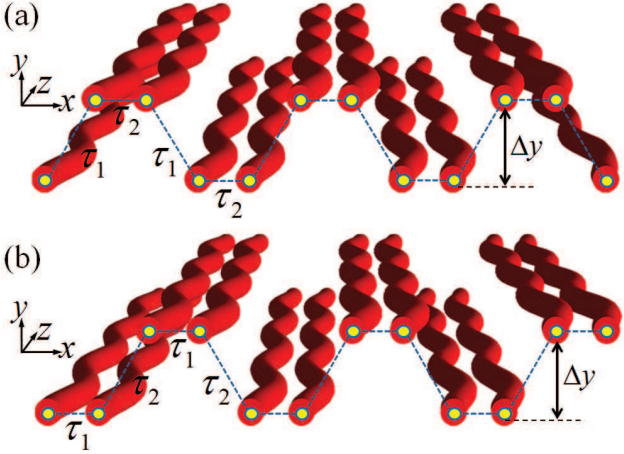}
  \caption{Schematic diagram of waveguide arrays curved along the propagation direction of light ($z$-axis). The center-to-center spacing along the x-axis is fixed as $\Delta x$, and the one along the y-axis is either 0 or $\Delta y$ intermittently. The coupling strength is either $\tau_1$ or $\tau_2$ intermittently. (a) $\tau_1/\tau_2<1$ with $\tau_2=\tau$, and (b) $\tau_1/\tau_2>1$ with $\tau_1=\tau$.} \label{fig1}
\end{figure}

We consider an array of coupled optical waveguides, where the waveguides are periodically curved along the longitudinal propagation direction, see Fig.~\ref{fig1}. The light field $\psi(x,y,z)$ obeys the paraxial wave equation
\begin{eqnarray} \label{PSequaion}
-i\frac{\partial \psi}{\partial z}&=&\frac{\lambda'}{4 \pi n'}\Big( \frac{\partial^2}{\partial x^2}
+\frac{\partial^2}{\partial y^2}\Big) \psi+\frac{2 \pi}{\lambda'}\nu(x,y,z)\psi,
\end{eqnarray}
where $\lambda'$ is the optical wavelength in vacuum, $n'$ is the medium refractive index, and $\nu(x,y,z)$ describes the refractive index at $(x,y,z)$. The waveguide centers $x_n(z)=x_n(z+T)$ are periodically curved along the longitudinal direction with the curving period $T$ much larger than the inter-waveguide distance $\Delta x$.
Here we set $x_n(z) = n\Delta x + A[\cos(\omega z)-1]$ with the modulation amplitude $A$ and the modulation frequency $\omega$.

By implementing the coordinate transformation: [$\hat{z}=z$, $\hat{y}=y$, $\hat{x}(z)=x-x_0(z)$], we have $\partial_x=\partial_{\hat{x}}$, $\partial_y=\partial_{\hat{y}}$ and $\partial_z=\partial_{\hat{z}}-\dot{x}_0\partial_{\hat{x}}$.
Therefore the field $\psi(\hat{x},\hat{y},\hat{z})$ obeys
\begin{eqnarray} \label{equation2}
-i\frac{\partial \psi}{\partial \hat{z}}&=&-i\dot{x}_0\frac{\partial \psi}{\partial \hat{x}}
+\frac{2 \pi}{\lambda'}\nu\psi
+\frac{\lambda'}{4 \pi n'}\Big (\frac{\partial^2 }{\partial \hat{x}^2}
+ \frac{\partial^2 }{\partial \hat{y}^2}\Big)\psi. \nonumber
\end{eqnarray}
By applying the gauge transformation
\begin{eqnarray} \label{equation3}
\psi=\phi\exp\{i\frac{\pi n'}{\lambda'}(2\dot{x}_0(\hat{z})\hat{x}(\hat{z})-\int^{\hat{z}}_0\hat{x}^2_0(\xi)d\xi)\}, \nonumber
\end{eqnarray}
the paraxial wave equation~(\ref{PSequaion}) can be written as
\begin{eqnarray} \label{equation4}
-i\frac{\partial \phi}{\partial \hat{z}}=\frac{\lambda'}{4 \pi n'}\Big ( \frac{\partial^2 }{\partial \hat{x}^2}+ \frac{\partial^2 }{\partial \hat{y}^2}\Big)\phi+\frac{2\pi}{\lambda'}\nu\phi-\frac{2 \pi n'}{\lambda'}\ddot{x}_0\hat{x}\phi. \nonumber
\end{eqnarray}
Expanding the field into a superposition of the single-mode fields in individual waveguides
$$
\phi(\hat{x},\hat{y},\hat{z})=\sum_{n}\varphi_n(\hat{z})a_n(\hat{x},\hat{y}),
$$
we obtain the coupled-mode equations
\begin{eqnarray} \label{equation5}
-i\frac{d\varphi_n}{d z}=\tau_n\varphi_{n+1}+\tau_{n-1}\varphi_{n-1}
+D_{n}\varphi_{n}-\eta\ddot{x}_0n\varphi_{n}, \nonumber
\end{eqnarray}
where $\eta=2\pi n' /\lambda'$ as a normalized optical frequency, and
\begin{eqnarray} \label{}
\tau_n&=&\frac{2\pi}{\lambda'}\int\int a_n^*(\hat{x},\hat{y})\nu(\hat{x},\hat{y},\hat{z})a_{n+1}(\hat{x},\hat{y})d\hat{x}d\hat{y},  \nonumber \\
D_{n}&=&\frac{2\pi}{\lambda'}\int\int a_n^*(\hat{x},\hat{y})\nu(\hat{x},\hat{y},\hat{z})a_{n}(\hat{x},\hat{y})d\hat{x}d\hat{y}.  \nonumber
\end{eqnarray}
By performing a transformation
\begin{eqnarray} \label{equation6}
\varphi_n=\exp[i\eta A \omega \hat{x}_n \sin(\omega z)+iD_{n}z]u_n, \nonumber
\end{eqnarray}
we derive the coupled-mode equations become as
\begin{eqnarray} \label{equation8}
-i\frac{du_{n}}{d z}&=&\tau_n \exp[i\eta A \omega (\hat{x}_{n+1}-\hat{x}_{n}) \sin(\omega z)]u_{n+1} \\
&&+\tau_{n-1} \exp[-i\eta A \omega (\hat{x}_{n}-\hat{x}_{n-1}) \sin(\omega z)]u_{n-1}. \nonumber
\end{eqnarray}
Here $\eta=2\pi n'/\lambda'$, $u_n$ denotes the complex field amplitude for the $n$-th waveguide with $n$ being the waveguide index.
As the center-to-center waveguide spacing along the x-axis is constant (i.e. $\hat{x}_{n+1}-\hat{x}_n=\hat{x}_{n}-\hat{x}_{n-1}=\Delta \hat{x}=1$) and one along the y-axis is either 0 or $\Delta y$ intermittently, the hopping strengths can be written as $\tau_{n}=\frac{1}{2}\{[1-(-1)^n]\tau_1+[1+(-1)^n]\tau_2 \}$ and the maximum hopping strength $\tau=\max\{\tau_1,\tau_2\}$ is fixed.
By adjusting the distance $\Delta \hat{y}$, one may tune the values of $\tau_{n}$.

Without loss of generality, we set $\eta=1$ and $\tau=1$.
Therefore the system can be described by the periodically modulated SSH-like Hamiltonian
\begin{eqnarray} \label{hmd}
H(z)=\sum\limits_{n=1}^{2N}(\tau_{n} \exp[i A \omega \sin(\omega z)] u_n^* u_{n+1}+h.c.),
\end{eqnarray}
with $2N$ being the total number of optical waveguides.
Chiral symmetry is represented by the sublattice operator $\Gamma=\sum\limits_{n}^N u_{2n-1}^*u_{2n-1}-\sum\limits_{n}^N u_{2n}^*u_{2n}$, which is unitary, Hermitian and local.
Obviously, $\Gamma H \Gamma=-H$, this means that this periodically modulated SSH-like Hamiltonian has chiral symmetry~\cite{asboth2016short}.
On the other hand, the above Hamiltonian also has time reversal symmetry, i.e. it is invariant under the transformation [$z\rightarrow -z, i\rightarrow -i$].

\section{Floquet Energy Spectrum\label{Secss}}
Since the system is invariant under $z\rightarrow z+T$, according to the Floquet theorem~\cite{Gomez-Leon:2013-200403:PRL}, the steady states of the coupled-mode equation~(\ref{equation8}) follow
\begin{eqnarray}
{u_n}(z) = {e^{ - i E z}}\sum\limits_{\chi  =  - \infty}^{+ \infty} {{e^{ - i\chi \omega z}}{c_{n,\chi }}}, \nonumber
\end{eqnarray}
where $c_{n,\chi }$  is the amplitude of the $\chi$-th Floquet state.
Substituting the above Floquet expansion into the coupled-mode equations, one obtain quasi-energy equation in the Floquet space
\begin{eqnarray}
E {c_{n,\chi }} &=& \sum\limits_{\chi ' =  - \infty}^{+ \infty} {{\tau _{n - 1}}{e^{-i\eta A \omega \sin(\omega z)}e^{-i(\chi'-\chi)\omega z}}{c_{n - 1,\chi '}}} \nonumber \\
&&+\sum\limits_{\chi' = -\infty}^{+ \infty} {{\tau _n}{e^{i\eta A \omega \sin(\omega z)}e^{-i(\chi'-\chi)\omega z}}{c_{n{\rm{ + }}1,\chi '}}} \nonumber \\
&&+\sum\limits_{\chi' = -\infty}^{+ \infty} \chi' \omega e^{-i(\chi'-\chi)\omega z} {c_{n,\chi' }}\nonumber \\
&&+\sum\limits_{\chi'\neq\chi}e^{-i(\chi'-\chi)\omega z}E {c_{n,\chi' }}. \nonumber
\end{eqnarray}
We introduce the average over one modulation period for all $z$-dependent quantities and obtain the quasi-energy eigen mode equation
\begin{eqnarray}
E {c_{n,\chi }} &=& \sum\limits_{\chi ' =  - \infty}^{+ \infty} {{\tau _{n - 1}}{J_{\chi  - \chi '}}{c_{n - 1,\chi '}}} \nonumber \\
&&+\sum\limits_{\chi' = -\infty}^{+ \infty} {{\tau _n}{J_{\chi ' - \chi }}{c_{n{\rm{ + }}1,\chi '}}} +\chi \omega {c_{n,\chi }},
\end{eqnarray}
where $J_{\chi'-\chi}$ is the Bessel function $J_{\chi'-\chi}(A \omega)$.
To obtain the quasi-energy spectrum, one needs to truncate the Floquet space.
In our calculation, we choose $\chi',\chi\in [-X,X] $ and $Y=2X+1$ is the truncation number.

\begin{figure}[htp]
\center
\includegraphics[width=\columnwidth]{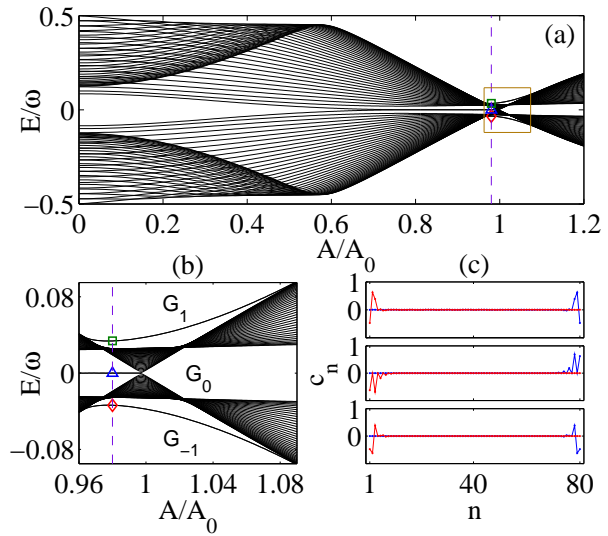}
  \caption{Quasi-energy spectra under open boundary condition. (a) Scaled quasi-energy $E/\omega$ vs. the scaled modulation amplitude $A/A_0$. (b) Enlarged rectangular region of (a). (c) The Floquet edge states corresponding to the square, triangle, and diamond points in the three gaps $(G_{+1}, G_0, G_{-1})$ at $A/A_0=0.98$ marked in (b). The parameters are chosen as $\tau_1/\tau_2=1.2$, $2\pi/\omega=3$, $A_0\omega\simeq 2.405$ [which gives $J_0(A_0 \omega)=0$], the total lattice number $2N=80$ and the truncation number $Y=13$.} \label{fig2}
\end{figure}

Now we discuss the quasi-energy spectra under OBC.
In Figs.~\ref{fig2}(a) and~\ref{fig2}(b), we show the scaled quasi-energy $E/\omega$ versus the scaled modulation amplitude $A/A_0$.
In our calculation, $A_0$ is given by the first zero-point of $J_0(A_0 \omega)$, $\tau_1/\tau_2=1.2$, $2\pi/\omega=3$, and the total lattice number $2N=80$.
In the energy gap $G_0$, there appear isolated zero-energy levels under some parameters ranges.
Because the quasi-energies have periodicity in Floquet space, so that similar isolated levels can also appear in gaps $G_{\pm 2, \pm 4, ...}$.
In the energy gaps $G_{-1}$ and $G_1$, isolated nonzero-energy levels appear around $A/A_0\sim 1$ and the similar isolated levels can also appear in gap $G_{\pm 3, \pm 5, \cdots}$.
Below, we concentrate our discussion on the quasi-energy ranges $-1/2\leq E/\omega \leq 1/2$.
In particular, isolated zero- and nonzero-energy levels can coexist in the same parametric region, see Fig.~\ref{fig2}(b).
The eigenstate profiles, which localize at two edges, indicate that these isolated levels are FESs [see Fig.~\ref{fig2}(c)].
We know that the topological edge states in a static SSH model always appear as zero-energy modes.
However, in our modulated system, there appear both zero- and nonzero-energy edge states.
Naturally, there arises an open question: \emph{Are all FESs induced by topology?}

\section{Multi-Scale Analysis\label{Sec3}}

To understand how FESs appear in the high-frequency limit, we employ the multi-scale perturbation analysis~\cite{Kivshar:1994-2536:PRE, Garanovich:2008-203904:PRL}.
We rewrite Eq.(\ref{equation8}) as
\begin{eqnarray} \label{equation20s}
-i\frac{du_n}{dz}=\sum\limits_m W(z;n,m)u_m,
\end{eqnarray}
with
\begin{eqnarray} \label{equation21s}
W(z;n,m)&=&\frac{1+(-1)^n}{2}[\delta_{n,m+1}\tau_1e^{-i A \omega \sin(\omega z)} \nonumber \\
&&+\delta_{n,m-1}\tau_2e^{i A \omega \sin(\omega z)}]  \nonumber \\
&&+\frac{1-(-1)^n}{2}[\delta_{n,m+1}\tau_2e^{-i A \omega \sin(\omega z)} \nonumber \\
&&+\delta_{n,m-1}\tau_1e^{i A \omega \sin(\omega z)}].  \nonumber
\end{eqnarray}
For the open boundary condition, we have $u_{n<1}\equiv 0$ and $u_{n>2N}\equiv 0$, in which $2N$ is the total lattice number.
Therefore, $W(z;n,m)$ can be rewritten as
\begin{eqnarray} \label{equation22s}
W(z;n,m)&=&\frac{1+(-1)^n}{2}[\delta_{n,m+1}\tau_1e^{-i A \omega \sin(\omega z)}\nonumber \\
&&+(1-\delta_{n,2N})\delta_{n,m-1}\tau_2e^{i A \omega \sin(\omega z)}] \nonumber \\
&&+\frac{1-(-1)^n}{2}[(1-\delta_{n,1})\delta_{n,m+1}\tau_2e^{-i A \omega \sin(\omega z)} \nonumber \\
&&+\delta_{n,m-1}\tau_1e^{i A \omega \sin(\omega z)}].
\end{eqnarray}
Because the waveguide axes are periodically curved along the longitudinal propagation ($z$-direction), we have $W(z;n,m)=W(z+T;n,m)$, where $T=2\pi/\omega$.
In the high-frequency limit ($\omega \gg 1$), we can introduce a small parameter $\varepsilon$, which satisfy $T=O(\varepsilon)$.
Thus, the solution of Eq.(\ref{equation20s}) can be given as the series expansion
\begin{eqnarray} \label{equation23s}
u_n(z)&=&U_n(z_0,z_1,z_2,...)+\varepsilon v_n(z_{-1},z_0,z_1,z_2,...) \nonumber \\
&&+\varepsilon^2 w_n(z_{-1},z_0,z_1,z_2,...) \nonumber \\
&&+\varepsilon^3 \zeta_n(z_{-1},z_0,z_1,z_2,...)+O(\varepsilon^4),
\end{eqnarray}
where $z_{l'}=\varepsilon^{l'} z$.
Then the differentiation is performed according to the usual convention:
\begin{eqnarray} \label{equation24s}
\frac{d}{dz}=\varepsilon^{-1}\frac{\partial}{\partial z_{-1}}+\frac{\partial}{\partial z_0}+
\varepsilon\frac{\partial}{\partial z_1}+\varepsilon^2\frac{\partial}{\partial z_2}+\cdots.
\end{eqnarray}
In the series solution, the function $U_n$ describes the averaged behavior
\begin{eqnarray} \label{equation25s}
\langle u_n \rangle=U_n ;
\langle \frac{du_n}{dz} \rangle=\frac{dU_n}{dz},
\end{eqnarray}
in which the average notation
\begin{eqnarray} \label{equation26s}
\langle \bullet \rangle =\varepsilon T^{-1}\int_{\varepsilon^{-1} z}^{\varepsilon^{-1} (z+T)}(\bullet)(z_{-1})dz_{-1}. \nonumber
\end{eqnarray}
It is worth to note that $U_n$ does not depend on the `fast' variable $z_{-1}$, this means that
\begin{eqnarray} \label{equation27s}
\langle U_n \rangle=U_n ;
\langle \frac{dU_n}{dz} \rangle=\frac{dU_n}{dz}.
\end{eqnarray}
From Eqs.(\ref{equation25s}) and (\ref{equation27s}), we have
\begin{eqnarray} \label{equation28s}
&&\langle v_n \rangle=\langle w_n \rangle=\langle \zeta_n \rangle \equiv0 ; \nonumber \\
&&\langle \frac{\partial v_n}{\partial z_{l'}} \rangle=\langle \frac{\partial w_n}{\partial z_{l'}} \rangle=\langle \frac{\partial \zeta_n}{\partial z_{l'}} \rangle \equiv0,
\end{eqnarray}
for $l'=-1,0,1,2,\cdots$.

Substituting Eq.(\ref{equation23s}) into Eq.(\ref{equation20s}) and collecting terms with different orders of $\varepsilon$, we obtain
\begin{eqnarray} \label{equation29s}
-i\frac{\partial U_n}{\partial z_0}=i\frac{\partial v_n}{\partial z_{-1}}+\sum\limits_m W(z;n,m)U_m,
\end{eqnarray}
for the order $\varepsilon^0$.
Using the conditions Eq.(\ref{equation27s}) and Eq.(\ref{equation28s}) and averaging Eq.(\ref{equation29s}), we have
\begin{eqnarray} \label{equation30s}
-i\frac{\partial U_n}{\partial z_0}=\sum\limits_m W_0(n,m)U_m,
\end{eqnarray}
where
$W_0(n,m)=\langle W(z;n,m) \rangle$.
Then substituting Eq.(\ref{equation30s}) into Eq.(\ref{equation29s}), we can obtain the equation for $v_n$
\begin{eqnarray} \label{equation32s}
-i\frac{\partial v_n}{\partial z_{-1}}=\sum\limits_m [W(z;n,m)-W_0(n,m)]U_m.
\end{eqnarray}
Thus through integrating the above equation, we derive an explicit expression for the function $v_n$
\begin{eqnarray} \label{equation33s}
v_n=i\varepsilon^{-1}\sum\limits_m M(z;n,m)U_m ,
\end{eqnarray}
with
$M(z;n,m)=\int[W(z;n,m)-W_0(n,m)]dz$.
Here, the function $M$ is periodic and has average zero value
\begin{eqnarray} \label{equation35s}
M(z;n,m)\equiv M(z+T;n,m) ; \langle M(z;n,m) \rangle=0.
\end{eqnarray}

For the order $\varepsilon^1$, we have
\begin{eqnarray} \label{equation36s}
-i\frac{\partial U_n}{\partial z_1}=i\frac{\partial v_n}{\partial z_0}+i\frac{\partial w_n}{\partial z_{-1}}+\sum\limits_m W(z;n,m)v_m.
\end{eqnarray}
Substituting Eqs.(\ref{equation33s}) and (\ref{equation30s}) into Eq.(\ref{equation36s}), we obtain
\begin{eqnarray} \label{equation37s}
-i\frac{\partial U_n}{\partial z_1}&=&-i\varepsilon^{-1}\sum\limits_{m,j} M(z;n,j)W_0(j,m)U_m+i\frac{\partial w_n}{\partial z_{-1}}  \nonumber \\
&&+i\varepsilon^{-1}\sum\limits_{m,j}W(z;n,j)M(z;j,m)U_m.
\end{eqnarray}
Using the conditions (\ref{equation27s}), (\ref{equation28s})  and (\ref{equation35s}) and averaging Eq.(\ref{equation37s}), we have
\begin{eqnarray} \label{equation38s}
-i\frac{\partial U_n}{\partial z_1}=i\varepsilon^{-1}\sum\limits_{m,j}\langle W(z;n,j)M(z;j,m) \rangle U_m.
\end{eqnarray}
Substituting Eq.(\ref{equation38s}) into Eq.(\ref{equation37s}), we can obtain the equation for $w_n$
\begin{eqnarray} \label{equation39s}
-i\frac{\partial w_n}{\partial z_{-1}}&=&-i\varepsilon^{-1} \sum\limits_{m,j}M(z;n,j)W_0(j,m)U_m  \nonumber \\
&&+i\varepsilon^{-1} \sum\limits_{m,j}[W(z;n,j)M(z;j,m) \nonumber \\
&&-\langle W(z;n,j)M(z;j,m) \rangle]U_m.
\end{eqnarray}
Similarly, by performing integration, we can derive the explicit expression for $w_n$.

For the order $\varepsilon^2$, we have
\begin{eqnarray} \label{equation40s}
-i\frac{\partial U_n}{\partial z_2}&=&i\frac{\partial v_n}{\partial z_1}+i\frac{\partial w_n}{\partial z_0}+i\frac{\partial \zeta_n}{\partial z_{-1}} \nonumber \\
&&+\sum\limits_m W(z;n,m)w_m.
\end{eqnarray}
Using Eqs.(\ref{equation27s}) and (\ref{equation28s}) and averaging Eq.(\ref{equation40s}), we obtain
\begin{eqnarray} \label{equation41s}
-i\frac{\partial U_n}{\partial z_2}=\sum\limits_q \langle W(z;n,q)w_q \rangle,
\end{eqnarray}
where the second term
\begin{eqnarray} \label{equation42s}
\langle W(z;n,q)w_q \rangle&=&\langle [W(z;n,q)-W_0(n,q)] w_q \rangle \nonumber \\
&=& -\varepsilon^{-1} \langle M(z;n,q)\frac{\partial w_q}{\partial_{-1}} \rangle .
\end{eqnarray}
Then using Eqs.(\ref{equation39s}) and (\ref{equation28s}), we can rewrite Eq.(\ref{equation41s}) as
\begin{eqnarray} \label{equation43s}
-i\frac{\partial U_n}{\partial z_2}&=&\varepsilon^{-2}\sum\limits_{q,m,j}\langle M(z;n,q) [W(z;q,j) \nonumber \\
&&-W_0(q,j)] M(z:j,m) \rangle U_m \nonumber \\
&&+\varepsilon^{-2}\sum\limits_{q,m,j} \langle M(z;n,q)[W_0(q,j)M(Z;j,m) \nonumber \\
&&-M(z;q,j)W_0(j,m)] \rangle U_m.
\end{eqnarray}

By combining Eqs.(\ref{equation30s}), (\ref{equation38s}) and (\ref{equation43s}) and using Eq.~(\ref{equation24s}), we obtain a closed-form equation for $U_n$
\begin{eqnarray} \label{equation44s}
-i\frac{d U_n}{dz}=\sum\limits_{m} W_s(n,m) U_m .
\end{eqnarray}
Here the effective coupling coefficients are given as
\begin{eqnarray} \label{equation45s}
W_s(n,m)&=&W_0(n,m)+\sum\limits_j W_1(n,j,m) \nonumber \\
&&+\sum\limits_{q,j}W_2(n,q,j,m),
\end{eqnarray}
with
\begin{eqnarray} \label{}
W_0(n,m)&=&\langle W(z;n,m) \rangle=\frac{1+(-1)^n}{2}[\delta_{n,m+1}\tau_1 \nonumber \\
&&+(1-\delta_{n,2N})\delta_{n,m-1}\tau_2]J_0(\eta A \omega) \nonumber \\
&&+\frac{1-(-1)^n}{2} [(1-\delta_{n,1})\delta_{n,m+1}\tau_2 \nonumber \\
&&+\delta_{n,m-1}\tau_1] J_0(\eta A \omega), \nonumber
\end{eqnarray}
\begin{eqnarray} \label{}
\sum\limits_j W_1(n,j,m)=i\sum\limits_j \langle W(z;n,j)M(z;j,m) \rangle=0, \nonumber
\end{eqnarray}
\begin{eqnarray} \label{}
&&\sum\limits_{q,j} W_2(n,q,j,m) \nonumber \\
&=&\sum\limits_{q,j} \langle M(z;n,q)[W(z;q,j) \nonumber \\
&&-W_0(q,j)]M(z;j,m) \rangle \nonumber \\
&&+\sum\limits_{q,j} \langle M(z;n,q)[W_0(q,j)M(z;j,m) \nonumber \\
&&-M(z;q,j)W_0(j,m)] \rangle \nonumber \\
&=&\frac{1+(-1)^n}{2}\{\delta_{n,m+1}[(\tau_1/\tau_2)^2-1](\tau_1/\tau_2) \nonumber \\
&&+\delta_{n,m-1}[1-(\tau_1/\tau_2)^2]\}\Delta \nonumber \\
&&+\frac{1-(-1)^n}{2}\{\delta_{n,m+1}[1-(\tau_1/\tau_2)^2] \nonumber \\
&&+\delta_{n,m-1}[(\tau_1/\tau_2)^2-1](\tau_1/\tau_2)\}\Delta \nonumber \\
&&+\frac{\tau_1}{2 \tau_2}(\delta_{n,1}\delta_{m,2}+\delta_{n,2}\delta_{m,1} \nonumber \\
&&+\delta_{n,2N}\delta_{m,2N-1}+\delta_{n,2N-1}\delta_{m,2N}) \Delta, \nonumber
\end{eqnarray}
with
\begin{eqnarray} \label{equation46s}
\Delta&=&-\omega^{-2}\tau_2^3 \sum\limits_{m\neq0}\sum\limits_{j\neq0,-m}J_j(A \omega)  \nonumber \\
&&J_m(A \omega)J_{j+m}(A \omega)j^{-1}m^{-1}. \nonumber
\end{eqnarray}
Finally, the effective equations for the slowly varying functions $U_n(z)$ read as
\begin{eqnarray} \label{equation47s}
-i\frac{dU_{2n-1}}{dz}&=&\tau_a U_{2n}+\tau_b U_{2n-2}+\delta_{(2n-1,1)}\tau_c U_2 \nonumber \\
&&+\delta_{(2n-1,2N-1)}\tau_c U_{2N},  \nonumber \\
-i\frac{dU_{2n}}{dz}&=&\tau_b U_{2n+1}+\tau_a U_{2n-1}+\delta_{(2n,2)}\tau_c U_1 \nonumber \\
&&+\delta_{(2n,2N)}\tau_c U_{2N-1}.
\end{eqnarray}
with the Kronecker's delta-function $\delta_{(n,m)}$.  
Here, the effective couplings are given as
\begin{eqnarray} \label{equation48s}
\tau_a&=&\tau_1 J_0-(\tau_1/\tau_2)\Theta, \nonumber  \\
\tau_b&=&\tau_2 J_0 +\Theta, \nonumber \\
\tau_c&=&\tau_1 \Delta / (2 \tau_2).
\end{eqnarray}
with $\Delta=-\omega^{-2}\tau_2^3 \sum\limits_{m\neq0}\sum\limits_{j\neq \{0,-m\}}J_j J_m J_{j+m}j^{-1}m^{-1}$ and $\Theta=[1-(\tau_1/\tau_2)^2]\Delta$.
The effective couplings $\tau_c$ describe the virtual defects at boundaries, as shown in the schematic diagram in Fig.~\ref{fig4}.

Based on the above discussions, the periodically modulated system can be description by an effective static SSH-like coupled-mode Eqs.~(\ref{equation47s}).
The major differences is the existence of virtual defects at boundaries in the effective model.
Similar to a surface perturbation, the virtual defects can form a defect-free surface states (or FESs)~\cite{Garanovich:2008-203904:PRL}.
On the other hand, if $\tau_c=0$, the static SSH-like coupled-mode equations reduce to conventional SSH model~\cite{Su:1979-1698:PRL} and the defect-free surface states disappear.
However, for the 1D conventional SSH model belongs to the BDI symmetry class~\cite{ryu2010topological},
which satisfy time reversal and chiral symmetry, can support an $Z$ topological index (the integer $Z$ index can only take values 0 or 1) \cite{ganeshan2013topological}.
For $|\tau_a|/|\tau_b|<1$, this system is topologically nontrivial and has one zero-energy mode localized at each edge, the zero-energy edge mode also call Shockley-like surface states~\cite{malkova2009transition}.
For $|\tau_a|/|\tau_b|>1$, the system is topologically trivial with no edge modes.
If change $\tau_c\neq0$, the static SSH-like coupled-mode equations still satisfy time reversal and chiral symmetry, which illustrate that the multi-scale perturbation analysis do not change the symmetry of the system.
In similarly static system, the relation between Shockley-like and Tamm-like surface states has been discussed~\cite{malkova2007interplay, malkova2009transition, malkova2009observation}.
Their results show that the transitions between Shockley-like and Tamm-like surface states are observed by tuning the surface perturbation (embedded defects).
In our system, without any embedded or nonlinearity-induced defects, the surface perturbation (virtual defects) is induced by periodical modulations.
In the next section, we will give the parameter regions of FESs and explore their topological nature.

\section{Non-Topological vs. Topological Edge States\label{Sec4}}

\subsection{Asymptotic phase boundary \label{A}}

To estimate the cutoff values (phase boundaries) for the regions of FESs caused by virtual defects.
We now consider stationary solutions in the form of $U_n(z)=U_n(0)e^{iEz}$ with $E$ being the propagation constant.
Substituting it into Eq.(\ref{equation47s}), we obtain
\begin{eqnarray} \label{equation49s}
E U_{2n-1}&=&\tau_a U_{2n}+\tau_b U_{2n-2} \nonumber \\
&&+(\delta_{2n-1,1}\tau_c U_2+\delta_{2n-1,2N-1}\tau_c U_{2N}) \nonumber \\
E U_{2n}&=&\tau_b U_{2n+1}+\tau_a U_{2n-1}\nonumber \\
&&+(\delta_{2n,2}\tau_c U_1+\delta_{2n,2N}\tau_c U_{2N-1}).
\end{eqnarray}
For an infinite lattice, we have
\begin{eqnarray} \label{equation50s}
E U_{2n-1}&=&\tau_a U_{2n}+\tau_b U_{2n-2}, \nonumber \\
E U_{2n}&=&\tau_b U_{2n+1}+\tau_a U_{2n-1}.
\end{eqnarray}
The solution of Eqs.(\ref{equation50s}) can be given as the ansatz
\begin{eqnarray} \label{equation51s}
U_{2n-1}&=&a_1 Qe^{ikn}+a_2 Pe^{-ikn}, \nonumber \\
U_{2n}&=&a_1 Pe^{ikn}+a_2 Qe^{-ikn},
\end{eqnarray}
where $a_1$ and $a_2$ are arbitrary nonzero constants.
Substituting Eqs.(\ref{equation51s}) into Eqs.(\ref{equation50s}), we obtain
\begin{eqnarray} \label{equation52s}
E \left[
\begin{matrix}
P\\
Q
\end{matrix}
\right]
=\left[\begin{matrix}
0&\tau_a+\tau_be^{ik}&\\
\tau_a+\tau_be^{-ik}&0&
\end{matrix}
\right]
\left[
\begin{matrix}
P\\
Q
\end{matrix}
\right],
\end{eqnarray}
then we can have
\begin{eqnarray} \label{equation52a}
\frac{P}{Q}=\frac{E}{\tau_a+\tau_b e^{-ik}}=\frac{\tau_a+\tau_b e^{ik}}{E}.
\end{eqnarray}
Therefore, the propagation constant is given as
\begin{eqnarray} \label{equation53s}
E^2=\tau_a^2+\tau_b^2+2\tau_a\tau_b \cos(k),
\end{eqnarray}
for $k\in[-\pi, \pi]$.

For a finite but sufficiently large number of lattices ($2N=80$ in our calculation), consider the two edges, we have
\begin{eqnarray} \label{equation54s}
EU_2&=&(\tau_a+\tau_c)U_1+\tau_bU_3, \nonumber \\
EU_1&=&(\tau_a+\tau_c)U_2, \nonumber \\
EU_{2N}&=&(\tau_a+\tau_c)U_{2N-1}, \nonumber \\
EU_{2N-1}&=&(\tau_a+\tau_c)U_{2N}+\tau_bU_{2N-2}.
\end{eqnarray}
Besides $U_1$ and $U_{2N}$, the coupling equations is consistent with the Eqs.(\ref{equation50s}).
So that we should rewrite the ansatz, similarly the Eqs.(\ref{equation51s}), we have
\begin{eqnarray} \label{equation51a}
U_{2n-1}&=&U_1 \qquad (n=1), \nonumber \\
U_{2n-1}&=&a_1 Qe^{ikn}+a_2 Pe^{-ikn} \quad (1<n\leq N), \nonumber \\
U_{2n}&=&a_1 Pe^{ikn}+a_2 Qe^{-ikn} \quad (1\leq n<N),  \nonumber \\
U_{2n}&=&U_{2N} \qquad (n=N).
\end{eqnarray}
First, we consider left boundary of lattices and we can give a set of equation
\begin{eqnarray} \label{equation51b}
EU_2&=&(\tau_a+\tau_c)U_1+\tau_bU_3, \nonumber \\
EU_1&=&(\tau_a+\tau_c)U_2, \nonumber \\
EU_{2(N-1)}&=&\tau_a U_{2(N-1)-1}+ \tau_b U_{2(N-1)+1}.
\end{eqnarray}
Combining Eqs.(\ref{equation51a}) and Eqs.(\ref{equation51b}), we have
\begin{eqnarray} \label{equation51c}
&&\frac{e^{-ik2(N-1)}}{e^{ik2(N-1)}}=  \nonumber \\
&&\frac{[\tau_b\frac{P}{Q}e^{-ik}+\frac{(\tau_a+\tau_c)^2}{E}-E](E\frac{P}{Q}-\tau_b e^{ik}-\tau_a e^{-ik})}{[E\frac{P}{Q}-\tau_b e^{ik}-\frac{(\tau_a+\tau_c)^2}{E}\frac{P}{Q}](\tau_b\frac{P}{Q}e^{-ik}-E+\tau_a\frac{P}{Q}e^{ik})}. \nonumber \\
\end{eqnarray}
We set $k=-i\varrho$ have $\frac{e^{-ik2(N-1)}}{e^{ik2(N-1)}}=e^{-4 \varrho (N-1)}$, where $\varrho$ is real number. If $\varrho>0$, when $N\rightarrow \infty$ have $e^{-4 \varrho (N-1)}\simeq 0$ and  equivalent to
\begin{eqnarray} \label{equation51d}
[\tau_b\frac{P}{Q}e^{-\varrho}+\frac{(\tau_a+\tau_c)^2}{E}-E](E\frac{P}{Q}-\tau_b e^{\varrho}-\tau_a e^{-\varrho})\simeq 0. \nonumber \\
\end{eqnarray}
If $\varrho<0$, when $N\rightarrow \infty$ have $e^{-4 \varrho (N-1)}\simeq \infty$ and  equivalent to
\begin{eqnarray} \label{equation551d}
[E\frac{P}{Q}-\tau_b e^{\varrho}-\frac{(\tau_a+\tau_c)^2}{E}\frac{P}{Q}](\tau_b\frac{P}{Q}e^{-\varrho}-E+\tau_a\frac{P}{Q}e^{\varrho}) \simeq 0. \nonumber \\
\end{eqnarray}
Combining Eq.(\ref{equation52a}) and Eq.(\ref{equation51d}), we have
\begin{eqnarray} \label{equation55s}
e^{\varrho}=\frac{\tau_c(\tau_c+2\tau_a)}{\tau_a\tau_b}=e^{ik}=d.
\end{eqnarray}
Similarly, Combining Eq.(\ref{equation52a}) and Eq.(\ref{equation551d}), we have
\begin{eqnarray} \label{equation55aa}
e^{-\varrho}=\frac{\tau_a\tau_b}{\tau_c(\tau_c+2\tau_a)}=e^{-ik}=d^{-1}.
\end{eqnarray}
Thus in the vicinity of the self-collimation point [$J_0(A_0\omega)=0$], as the couplings $(\tau_a, \tau_b)$ are very weak, the edge states induced by the virtual defects with the quasi-energies $E_s$ is given as
\begin{eqnarray} \label{equation56}
E_s^2&=&\tau_a^2+\tau_b^2+\tau_a\tau_b [e^{ik}+e^{-ik}]  \nonumber \\
&=&\tau_a^2+\tau_b^2+\tau_a\tau_b [d+d^{-1}].
\end{eqnarray}
On the other hand, when we consider the right boundary of lattices, we can also obtain the surface energy $E_s$ and which is agree with Eq.(\ref{equation56}).

Obviously, when $E_s^2>\max(E^2)$, FESs appear in the energy gaps $G_{-1}$ and $G_{1}$.
Otherwise, when $E_s^2<\min(E^2)$, FESs appear in the gap $G_{0}$.
Obviously, $\max(E^2)$ and $\min(E^2)$ are given by $|\cos(k)|= 1$.
From $\cos(k)=+1$, one can obtain the cutoffs values
\begin{eqnarray} \label{equation581}
{A_{cs}^{1,2}}/{A_0} \simeq 1-\frac{\tau_1\tilde{\tau_c} \pm F_a}{\tau_1\tau_2}.
\end{eqnarray}
From $\cos(k)=-1$, one can obtain the cutoffs values
\begin{eqnarray} \label{equation582}
{A_{cs}^{3,4}}/{A_0} \simeq 1-\frac{-\tau_1\tilde{\tau_c} \pm F_b}{\tau_1\tau_2}.
\end{eqnarray}
Here, $A_0$ is the first root of the Bessel function $J_0(A \omega)=0$,
$F_a=\sqrt{(\tau_1\tilde{\tau_c})^2+\tau_1\tau_2 M_{+}}$,
$F_b=\sqrt{(\tau_1\tilde{\tau_c})^2+\tau_1\tau_2 M_{-}}$,
$\tilde{\tau_c}=\frac{\tau_1}{2 \tau_2}\tilde{\Delta}$,
$M_{\pm}=\frac{\tau_1}{\tau_2} \left[1-(\frac{\tau_1}{\tau_2})^2\right]~\tilde{\Delta} \left\{\left[1-(\frac{\tau_1}{\tau_2})^2\right] \tilde{\Delta}\mp2\tilde{\tau_c}\right\}\pm(\tilde{\tau_c})^2$, and
$\tilde{\Delta} = \Delta|_{A\rightarrow A_0}$.
These cutoff values define the boundaries between the regions with and without FESs, see the dashed blue curves in Figs.~\ref{fig4}(a) and (b), which also call defect-free surface states~\cite{Garanovich:2008-203904:PRL}.
Since $F_b$ is a purely imaginary number for all $2\pi/\omega$ when $\tau_1/\tau_2=1.2$, in Fig.~\ref{fig4}(a), there are no cutoff values $A_{cs}^{3,4}/A_0$.
When $2\pi/\omega\rightarrow 0$, all cutoff values gradually converge into one point at $A/A_0=1$, and there are no FESs caused by the virtual defects.

On the other hand, as the effective model Eq.~\eqref{equation47s} is an SSH-like model, the system changes from topological to non-topological when the effective coupling are tuned from $|\tau_a|<|\tau_b|$ to $|\tau_a|>|\tau_b|$.
The effective couplings $(\tau_a, \tau_b)$ depend on the original couplings $(\tau_1, \tau_2)$ and the driving parameters $(A, \omega)$.
We show the effective coupling strengthes $\left(|\tau_a|, |\tau_b|\right)$ versus the scaled modulation amplitude $A/A_0$ for $2\pi/\omega=2$ and $\tau_1/\tau_2=1.2$, see the inset in Fig.~\ref{fig4}(a).
There appear two intersection points at $|\tau_a|=|\tau_b|$ when $A/A_0$ increases.
In the regions of $|\tau_a|<|\tau_b|$, topological FESs appear [the relevant cumulative phase being $\pi$], which also call Shockley-like surface states~\cite{malkova2009transition}.
The intersection points, where topological phase transition points occur, are given by
\begin{eqnarray} \label{equation59}
A_{ct}^{5,6}/A_0 \simeq 1+\frac{(1\pm \tau_1/\tau_2)^2 \tilde{\Delta}}{\tau_2},
\end{eqnarray}
see the dashed blue curves $5$ and $6$ in inset of Fig.~\ref{fig4}(b)
Similarly, when $2\pi/\omega\rightarrow 0$, these two curves also gradually converge into one point at $A/A_0=1$.
Thus, in the limit of $2\pi/\omega = 0$, the effective couplings vanish when $A/A_0=1$ and the modulation does not change the topological feature when $A/A_0$ is tuned through $A/A_0=1$.

\subsection{Zak phase \label{B}}

To distinguish topological and non-topological FESs, we calculate the bulk topological invariant, the Zak phase~\cite{Zak:1989-2747:PRL}.
Zak phase can be predict the existence (with the relevant cumulative phase being $\pi$) or absence (vanishing
cumulative phase) of topological FESs in specific gap.

For a modulated SSH system of $N$ cells (i.e. $2N$ lattices) under PBC, by implementing a Fourier transform
\begin{eqnarray} \label{equation50sjb}
{c_{2n - 1,\chi }} &=& \frac{1}{{\sqrt N }}\sum\limits_k {{e^{ik(2n - 1)}}{c_{1,k,\chi }}}, \nonumber \\
{c_{2n,\chi }} &=& \frac{1}{{\sqrt N }}\sum\limits_k {{e^{ik2n}}{c_{2,k,\chi }}},
\end{eqnarray}
we obtain the quasi-energy spectra and the eigenstates by diagonalizing the quasi-energy equation
\begin{equation} \label{equationEE2}
{E _l}\left( {\begin{array}{*{20}{c}}
{c_{1,k,\chi }^{(l)}}\\
{c_{2,k,\chi }^{(l)}}
\end{array}} \right) = \sum\limits_{\chi '} {\widehat {\cal R}(k)\left( {\begin{array}{*{20}{c}}
{c_{1,k,\chi '}^{(l)}}\\
{c_{2,k,\chi '}^{(l)}}
\end{array}} \right)} + \chi \omega \hat I\left( {\begin{array}{*{20}{c}}
{c_{1,k,\chi }^{(l)}}\\
{c_{2,k,\chi }^{(l)}}
\end{array}} \right), \nonumber
\end{equation}
with the $2\times 2$ unit matrix $\hat I$ and the matrix
\begin{equation}
\widehat {\cal R}(k) = \left( {\begin{array}{*{20}{c}}
0&{{P_F}(k)}\\
{{{\tilde P}_F}(k)}&0
\end{array}} \right). \nonumber
\end{equation}
Here, ${P_F} =  {\tau _1}{J_{\chi ' - \chi }}{e^{ik}} + {\tau _2}{J_{\chi  - \chi '}}{e^{ - ik}}$, ${{\tilde P}_F} =  {\tau _1}{J_{\chi  - \chi '}}{e^{ - ik}} + {\tau _2}{J_{\chi ' - \chi }}{e^{ik}}$, and $k$ denotes the quasimomentum.

\begin{figure}[htp]
\center
\includegraphics[width=3.6in]{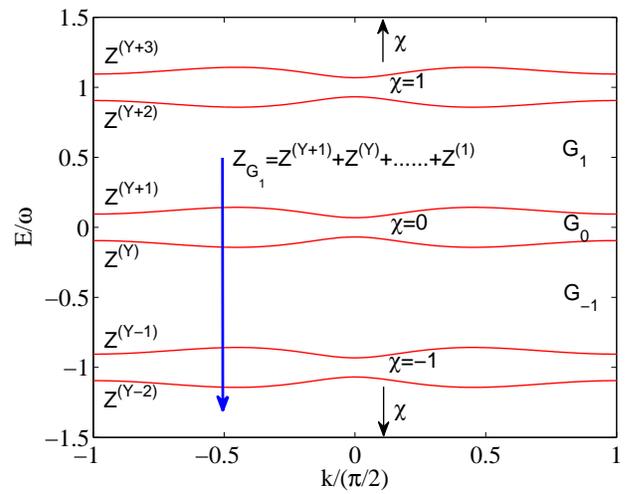}
\caption{Quasienergy spectrum in the quasi-momentum space and the Zak phase for the gap $G_1$.} \label{fig1ss}
\end{figure}

To compute the Zak phase for the Floquet quasi-energy spectrum one needs to truncate the Floquet space. The number of replicas needs to be chosen so that all relevant transitions at the desired energy are kept.
The Zak phase $Z_{G_m}$ for a specific gap is given by summing up $Z^{(l)}$ for all bands below the gap, where $Z^{(l)}=i\oint_k {\left\langle {c_k^{(l)}} \right|{\partial _k}\left| {c_k^{(l)}} \right\rangle dk}$ with the eigenstates ${\left| {c_k^{(l)}} \right\rangle }=\sum_{\alpha,\chi}c_{\alpha,k,\chi}^{(l)}\left|\alpha,k, \chi\right\rangle$ for the $l$-th band are superposition states of different Floquet-Bloch states $\left|\alpha,k,\chi\right\rangle$.
For a gap between the $(Y+m)$-th and $(Y+m+1)$-th bands, its Zak phase $Z_{G_m}$ is defined as
\begin{equation}
Z_{G_m}=\sum\limits_{l=1}^{Y+m}Z^{(l)}= \sum\limits_{l=1}^{Y+m}\big[i\oint_k {\left\langle {c_k^{(l)}} \right|{\partial _k}\left| {c_k^{(l)}} \right\rangle dk}\big].
\end{equation}
For example, the Zak phase $Z_{G_1}$ can be calculated by summing up all $Z^{(l)}$ for the bands below the gap $G_1$, see in Fig.~\ref{fig1ss}.

\subsection{Phase diagram \label{C}}

\begin{figure*}[htp]
\center
\includegraphics[width=1.3 \columnwidth]{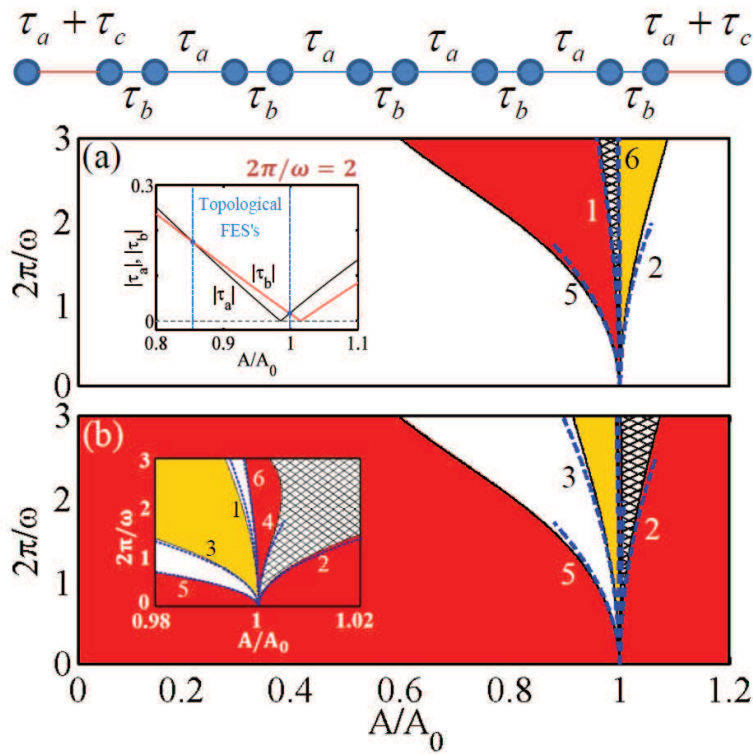}
  \caption{Phase diagram of the Floquet edge states.
  Top: Schematic diagram for the effective model Eq.~\eqref{equation47s}.
  (a,b)~Phase diagrams for (a) $\tau_1/\tau_2=1.2$ and (b) $\tau_2/\tau_1=1.2$.
  The red regions only support topological FESs, the yellow regions only support non-topological FESs, and the mesh regions support both topological and non-topological FESs.
  The curves $1$, $2$, $3$ and $4$ respectively correspond to the non-topological FESs cutoff values $A_{cs}^{1}/A_0$, $A_{cs}^{2}/A_0$, $A_{cs}^{3}/A_0$ and $A_{cs}^{4}/A_0$.
  While the curves $5$ and $6$ respectively correspond to the topological transition points $A_{ct}^{5}/A_0$ and $A_{ct}^{6}/A_0$, where the inset in (b) is the enlarged region nearby $A/A_0\sim0$.
  The system changes from topological to non-topological when the effective couplings are tuned from $|\tau_a|<|\tau_b|$ to $|\tau_a|>|\tau_b|$, see the inset in (a) for $2\pi/\omega=2$.
  } \label{fig4}
\end{figure*}

To verify the above analytical results, we numerically calculate the quasi-energy spectra.
From the quasi-energy spectra under OBC, we indeed find several FESs appear.
We then calculate Zak phases of the corresponding bulk states under PBC, find that the Zak phase $Z_{G_m}$ for a specific gap is either $0$ or $\pi$ and topological FESs only appear in a gap of nonzero $Z_{G_m}$.

In Fig.~\ref{fig4}, we show the phase diagram of all possible FESs in the parameter plane $(2\pi/\omega, A/A_0)$.
The appearance of topological FESs (red regions) and non-topological FESs (yellow regions) and their coexistence (mesh regions) sensitively depend on the coupling ratio $\tau_1/\tau_2$ and the modulation parameters $(\omega, A/A_0)$.
In the absence of modulation, topological edge states appear only if $\tau_1/\tau_2 < 1$, otherwise no edge state appears.
However, by applying a proper modulation, topological FESs may appear even if $\tau_1/\tau_2>1$ and also may disappear even if $\tau_1/\tau_2<1$.
In addition to the regions of topological and non-topological FESs, there exists the region of no edge states.
When $2\pi/\omega \rightarrow 0$, topological FESs appear if $\tau_1/\tau_2<1$ and all non-topological FESs gradually vanish at the zero-point of the Bessel function $J_0(A_0\omega)=0$.
Our numerical results clearly show all phase boundaries (the solid curves) gradually converge into one point at $A/A_0=1$ when $2\pi/\omega \rightarrow 0$, which well agree with our analytical results (the dashed blue curves).

\subsection{Non-coexistence of non-topological and topological Floquet edge states in the same gap \label{D}}

Although non-topological and topological FESs can be supported by the same parameters, we find that they can not appear in the same energy gap.
In this section, we only consider the quasi-energy ranges $ -1/2\leq E/\omega \leq 1/2$, so that the topological FESs (Zak Phase $Z_{G_0}=\pi$) only possible appear in gap $G_0$.
We will prove that non-topological and topological FESs can not coexist in the gap $G_0$.
For the whole Floquet spaces, due to the periodicity of quasi-energy.
This prove indirect reflection the topological FESs can not appear in the gap $G_{\pm 1, \pm 3, \pm 5,...}$, in addition the non-topological and topological FESs can not coexist in the gap $G_{0, \pm 2, \pm 4,...}$.

If non-topological FESs appear in the gap $G_0$, the edge state quasi-energy $E_s$ and the bulk-state quasi-energy $E$ will satisfy the condition $E_s^2<\min(E^2)$.
From Eq.~\eqref{equation53s}, the condition $E_s^2<\min(E^2)$ reads
\begin{eqnarray} \label{equation60}
E_s^2<\min(\tau_a^2+\tau_b^2-2\tau_a\tau_b, \tau_a^2+\tau_b^2+2\tau_a\tau_b),
\end{eqnarray}
which requests the parameters obeying $\tau_c(\tau_c+2\tau_a)<0$.
As the non-topological FESs appear around $A/A_0\sim 1$, we have $\tau_c=\frac{\tau_1}{2\tau_2}\Delta < 0$ and so that the above inequality is equivalent to
\begin{eqnarray} \label{equation62}
2\tau_1J_0(A\omega) -2\frac{\tau_1}{\tau_2}(1-(\frac{\tau_1}{\tau_2})^2)\Delta +\frac{1}{2}(\frac{\tau_1}{\tau_2})\Delta>0.
\end{eqnarray}
Below we separately discuss the two cases: (\textbf{I}) $\left.\tau_2>\tau_1>0\right.$ and (\textbf{II}) $\tau_1>\tau_2>0$.

\textbf{Case-I}: $\tau_2>\tau_1>0$. Without loss of generality, one can set $\tau_2=1$.

As $\tau_a<0$ always contradicts to the condition~\eqref{equation62}, the appearance of non-topological FESs in the gap $G_0$ requests
\begin{eqnarray} \label{equation63}
&&\tau_a=\tau_1 J_0(A\omega)-\tau_1(1-(\tau_1)^2)\Delta>0,  \nonumber \\
&&\tau_b=J_0(A\omega)+(1-(\tau_1)^2)\Delta<0,  \nonumber \\
&&2\tau_1J_0(A\omega) -2\tau_1(1-(\tau_1)^2)\Delta+\frac{1}{2}\tau_1\Delta>0,
\end{eqnarray}
or
\begin{eqnarray} \label{equation632}
&&\tau_a=\tau_1 J_0(A\omega)-\tau_1(1-(\tau_1)^2)\Delta>0,  \nonumber \\
&&\tau_b=J_0(A\omega)+(1-(\tau_1)^2)\Delta>0,  \nonumber \\
&&2\tau_1J_0(A\omega) -2\tau_1(1-(\tau_1)^2)\Delta+\frac{1}{2}\tau_1\Delta>0.
\end{eqnarray}

On the other hand, in the vicinity of $A_0$, we have $J_0(A \omega)<0$ when $A\rightarrow A_0^+$ and $J_0(A \omega)>0$ when $A\rightarrow A_0^-$.
Therefore, from the condition~\eqref{equation63}, one can obtain: (C1) $(0<\tau_1<\sqrt{1-\digamma}) \cap (0<\tau_1<\sqrt{3/4+\digamma})$ when $A\rightarrow A_0^-$, and (C2) $(0<\tau_1<\sqrt{3/4+\digamma})$ when $A\rightarrow A_0^+$.
Here, the parameter $\digamma$ is given as $\digamma=\frac{\omega^2 J_0(A\omega)}{4J_1^2(A\omega)J_2(A\omega)}$.
However, under the condition (C2), one can find that $E_s^2< 0$, this means that the condition (C2) does not support non-topological FESs in the gap $G_0$.
As we always have $\tau_b <0$ when $A\rightarrow A_0^+$, from the condition~\eqref{equation632}, we drive the condition (C3): $(\sqrt{1-\digamma}<\tau_1<\sqrt{3/4+\digamma})$ when $A\rightarrow A_0^-$.
Therefore, the appearance of non-topological FESs in the gap $G_0$ always request $A\rightarrow A_0^-$ (where $\tau_a>0$).

As the effective model \eqref{equation47s} is a SSH-like model, under the condition of $|\tau_a|/|\tau_b|<1$, the topological FESs are zero-energy modes and always appear in the gap $G_0$.
When $A\rightarrow A_0^-$ (where $\tau_a>0$), from $|\tau_a|/|\tau_b|<1$, one can obtain: (D1) $(0<\tau_1<1-\sqrt{\digamma})$ for $\tau_b<0$ and (D2) $(0<\tau_1<\sqrt{\digamma}-1)$ for $\tau_b>0$.
However, under the conditions (D1) and (D2), one can find that $E_s^2<0$, which means the absence of non-topological FESs.
That is to say, the non-topological and topological FESs can not coexist in the gap $G_0$.

\textbf{Case-II}: $\tau_1>\tau_2>0$. Without loss of generality, one can set $\tau_1=1$.

As $\tau_a<0$ always contradicts to the condition ~\eqref{equation62}, the existence of non-topological FESs in the gap $G_0$ requests $\tau_a>0$.
On the other hand, in the vicinity of $A_0$, we have $J_0(A \omega)<0$ when $A\rightarrow A_0^+$ and $J_0(A \omega)>0$ when $A\rightarrow A_0^-$.
When $\tau_1>\tau_2>0$, we always have $\tau_a<0$ when $A\rightarrow A_0^+$, so that the appearance of non-topological FESs in the gap $G_0$ always requests $A\rightarrow A_0^{-}$.
Moreover, when $A\rightarrow A_0^{-}$, we always have $\tau_b=\tau_2 J_0(A\omega)+(1-(\frac{1}{\tau_2})^2)\Delta>0$.
Thus the appearance of non-topological FESs in the gap $G_0$ requests
\begin{eqnarray} \label{equation64}
&&\tau_a=J_0(A\omega)-\frac{1}{\tau_2}(1-(\frac{1}{\tau_2})^2)\Delta>0,  \nonumber \\
&&\tau_b=\tau_2 J_0(A\omega)+(1-(\frac{1}{\tau_2})^2)\Delta>0,  \nonumber \\
&&2J_0(A\omega)-2\frac{1}{\tau_2}(1-(\frac{1}{\tau_2})^2)\Delta+\frac{1}{2}(\frac{1}{\tau_2})\Delta>0.
\end{eqnarray}
The above condition~\eqref{equation64} requests ($\frac{2 }{\sqrt{3}}\sqrt{1-\digamma}<\tau_2<1$).
However, under this condition, one can find that $E_s^2<0$, which means the absence of non-topological FESs.
This means that non-topological FESs can not appear in gap $G_0$ when $\tau_1>\tau_2>0$ and so that there is no coexistence of non-topological and topological FESs.

\begin{figure}[htp]
  \center
  \includegraphics[width=\columnwidth]{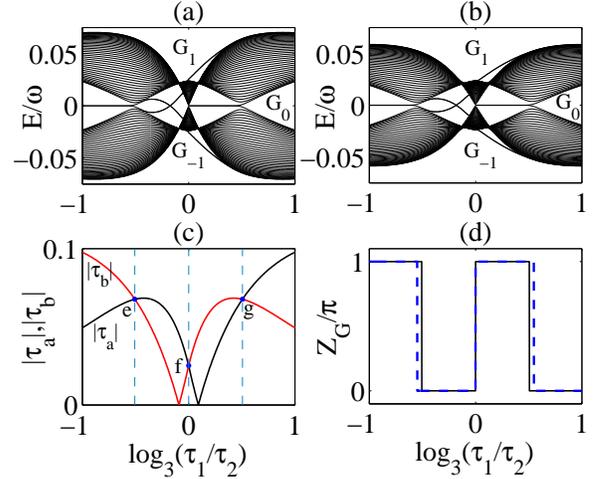}
  \caption{Scaled quasi-energy $E/\omega$ vs. coupling ratio $\tau_1/\tau_2$.
  (a) Band-gap structure of the effective model~\eqref{equation47s}.
  (b) Band-gap structure of the original model~\eqref{equation8}.
  (c) Effective coupling strengths ($|\tau_a|$, $|\tau_b|$) vs. the coupling ratio $\tau_1/\tau_2$.
  (d) The Zak phases for the gap $G_0$, in which the black and dashed blue lines correspond to the effective and original models, respectively.
  The parameters are chosen as $A/A_0=0.98$, $2\pi/\omega=3$, $A_0\omega\simeq 2.405$, the total lattice $2N=80$ and the truncation number $Y=13$.} \label{fig3}
\end{figure}

In order to explore how the ratio $\tau_1/\tau_2$ affects the FESs, we show how the scaled quasi-energy spectrum depends on $\tau_1/\tau_2$.
The quasi-energy spectra and Zak phases show that, even when the modulation frequency is not very high, the effective model may well explain the behaviors in the original system.
The deviation between the effective and original models decreases with the modulation frequency and gradually vanishes in the high-frequency limit.
In Fig.~\ref{fig3}, we show the quasi-energy spectra, the effective couplings and Zak phases for $A/A_0=0.98$ and $2\pi/\omega=3$.
Although the quasi-energies have small differences, the band-gap structures are almost the same, in which both zero and nonzero FESs may appear in different gaps, see Figs.~\ref{fig3}(a) and \ref{fig3}(b).
From the effective model, topological FESs are always zero-energy modes and only appear in the gap $G_0$ when $|\tau_a|/|\tau_b|<1$, see Figs.~\ref{fig3}(a) and \ref{fig3}(c).
In addition to the topological FESs, due to the modulation-induced virtual defects, there also exist non-topological FESs in different gaps.
Moreover, the band-gap structures show that topological and non-topological FESs can not appear in the same gap, which confirms our previous analytical analysis.
From the Zak phases, the effective and original models show similar topological phase transitions, but the transition points show small deviations dependent upon the modulation frequency, see Fig.~\ref{fig3}(d).

\section{Conclusion\label{Sec5}}

In summary, we have studied the Floquet edge states in arrays of curved optical waveguides described by the periodically modulated SSH model.
According to the Floquet theorem, we give the quasi-energy spectra under OBC and find several FESs.
To understand how FESs appear, we employ the multi-scale perturbation analysis and find the periodic modulations can induce virtual defects at boundaries.
Similar to a surface perturbation, the virtual defects can form a FESs (defect-free surface states)~\cite{Garanovich:2008-203904:PRL}.
On the other hand, by changing the ratio of $|\tau_a|/|\tau_b|$, one can also obtain a FESs (Shockley-like surface states).

In order to explore the topological nature of all FESs,
we have calculated the quasi-energy spectra and the Zak phases.
Our results indicate that Shockley-like surface state is a topological FES and defect-free surface state is a non-topological FES.
The topological and non-topological FESs may be supported by the same parameters, but they always appear in different energy gaps.
Without any embedded or nonlinearity-induced defects, these edge states originate from the interplay between the bulk band topology and periodic modulations.
We have derived analytically the boundaries between different topological phases, and have verified these results numerically.
We believe our work provides new perspectives for topological photonics govern by periodic modulations, and it can be employed for a control of topological phase transitions.
Although our analysis has been performed for arrays of periodically curved optical waveguides, it can be applicable to other lattice systems such as ultracold atoms in optical lattices~\cite{Atala:2013-795:NPHYS, Meier:2016-13986:NCOM}, photonic crystals~\cite{Zeuner:2015-40402:PRL}, and discrete quantum walks~\cite{Cardano:2017-15516:NCOM, Flurin:2017-31023:PRX}.

\begin{acknowledgments}
This work was supported by the National Natural Science Foundation of China (NNSFC) under Grants No. 11574405 and No. 11465008, and by the Australian Research Council. Both B. Zhu and H. Zhong made equal contributions.
\end{acknowledgments}

\bibliography{Floquet_edge_states}

\begin{thebibliography}{45}%
\makeatletter
\providecommand \@ifxundefined [1]{%
 \@ifx{#1\undefined}
}%
\providecommand \@ifnum [1]{%
 \ifnum #1\expandafter \@firstoftwo
 \else \expandafter \@secondoftwo
 \fi
}%
\providecommand \@ifx [1]{%
 \ifx #1\expandafter \@firstoftwo
 \else \expandafter \@secondoftwo
 \fi
}%
\providecommand \natexlab [1]{#1}%
\providecommand \enquote  [1]{``#1''}%
\providecommand \bibnamefont  [1]{#1}%
\providecommand \bibfnamefont [1]{#1}%
\providecommand \citenamefont [1]{#1}%
\providecommand \href@noop [0]{\@secondoftwo}%
\providecommand \href [0]{\begingroup \@sanitize@url \@href}%
\providecommand \@href[1]{\@@startlink{#1}\@@href}%
\providecommand \@@href[1]{\endgroup#1\@@endlink}%
\providecommand \@sanitize@url [0]{\catcode `\\12\catcode `\$12\catcode
  `\&12\catcode `\#12\catcode `\^12\catcode `\_12\catcode `\%12\relax}%
\providecommand \@@startlink[1]{}%
\providecommand \@@endlink[0]{}%
\providecommand \url  [0]{\begingroup\@sanitize@url \@url }%
\providecommand \@url [1]{\endgroup\@href {#1}{\urlprefix }}%
\providecommand \urlprefix  [0]{URL }%
\providecommand \Eprint [0]{\href }%
\providecommand \doibase [0]{http://dx.doi.org/}%
\providecommand \selectlanguage [0]{\@gobble}%
\providecommand \bibinfo  [0]{\@secondoftwo}%
\providecommand \bibfield  [0]{\@secondoftwo}%
\providecommand \translation [1]{[#1]}%
\providecommand \BibitemOpen [0]{}%
\providecommand \bibitemStop [0]{}%
\providecommand \bibitemNoStop [0]{.\EOS\space}%
\providecommand \EOS [0]{\spacefactor3000\relax}%
\providecommand \BibitemShut  [1]{\csname bibitem#1\endcsname}%
\let\auto@bib@innerbib\@empty
\bibitem [{\citenamefont {Lu}\ \emph {et~al.}(2014)\citenamefont {Lu},
  \citenamefont {Joannopoulos},\ and\ \citenamefont
  {Soljaclc}}]{Lu:2014-821:NPHOT}%
  \BibitemOpen
  \bibfield  {author} {\bibinfo {author} {\bibfnamefont {L.}~\bibnamefont
  {Lu}}, \bibinfo {author} {\bibfnamefont {J.~D.}\ \bibnamefont
  {Joannopoulos}}, \ and\ \bibinfo {author} {\bibfnamefont {M.}~\bibnamefont
  {Soljaclc}},\ }\bibfield  {title} {\enquote {\bibinfo {title} {Topological
  photonics},}\ }\href {\doibase 10.1038/NPHOTON.2014.248} {\bibfield
  {journal} {\bibinfo  {journal} {Nature Photonics}\ }\textbf {\bibinfo
  {volume} {8}},\ \bibinfo {pages} {821--829} (\bibinfo {year}
  {2014})}\BibitemShut {NoStop}%
\bibitem [{\citenamefont {Haldane}\ and\ \citenamefont
  {Raghu}(2008)}]{Haldane:2008-13904:PRL}%
  \BibitemOpen
  \bibfield  {author} {\bibinfo {author} {\bibfnamefont {F.~D.~M.}\
  \bibnamefont {Haldane}}\ and\ \bibinfo {author} {\bibfnamefont
  {S.}~\bibnamefont {Raghu}},\ }\bibfield  {title} {\enquote {\bibinfo {title}
  {Possible realization of directional optical waveguides in photonic crystals
  with broken time-reversal symmetry},}\ }\href {\doibase
  10.1103/PhysRevLett.100.013904} {\bibfield  {journal} {\bibinfo  {journal}
  {Phys. Rev. Lett.}\ }\textbf {\bibinfo {volume} {100}},\ \bibinfo {pages}
  {013904--4} (\bibinfo {year} {2008})}\BibitemShut {NoStop}%
\bibitem [{\citenamefont {Raghu}\ and\ \citenamefont
  {Haldane}(2008)}]{Raghu:2008-33834:PRA}%
  \BibitemOpen
  \bibfield  {author} {\bibinfo {author} {\bibfnamefont {S.}~\bibnamefont
  {Raghu}}\ and\ \bibinfo {author} {\bibfnamefont {F.~D.~M.}\ \bibnamefont
  {Haldane}},\ }\bibfield  {title} {\enquote {\bibinfo {title} {Analogs of
  quantum-{H}all-effect edge states in photonic crystals},}\ }\href {\doibase
  10.1103/PhysRevA.78.033834} {\bibfield  {journal} {\bibinfo  {journal} {Phys.
  Rev. A}\ }\textbf {\bibinfo {volume} {78}},\ \bibinfo {pages} {033834--21}
  (\bibinfo {year} {2008})}\BibitemShut {NoStop}%
\bibitem [{\citenamefont {Wang}\ \emph {et~al.}(2009)\citenamefont {Wang},
  \citenamefont {Chong}, \citenamefont {Joannopoulos},\ and\ \citenamefont
  {Soljacic}}]{Wang:2009-772:NAT}%
  \BibitemOpen
  \bibfield  {author} {\bibinfo {author} {\bibfnamefont {Z.}~\bibnamefont
  {Wang}}, \bibinfo {author} {\bibfnamefont {Y.~D.}\ \bibnamefont {Chong}},
  \bibinfo {author} {\bibfnamefont {J.~D.}\ \bibnamefont {Joannopoulos}}, \
  and\ \bibinfo {author} {\bibfnamefont {M.}~\bibnamefont {Soljacic}},\
  }\bibfield  {title} {\enquote {\bibinfo {title} {Observation of
  unidirectional backscattering-immune topological electromagnetic states},}\
  }\href {\doibase 10.1038/nature08293} {\bibfield  {journal} {\bibinfo
  {journal} {Nature}\ }\textbf {\bibinfo {volume} {461}},\ \bibinfo {pages}
  {772--U20} (\bibinfo {year} {2009})}\BibitemShut {NoStop}%
\bibitem [{\citenamefont {Umucalilar}\ and\ \citenamefont
  {Carusotto}(2011)}]{Umucalilar:2011-43804:PRA}%
  \BibitemOpen
  \bibfield  {author} {\bibinfo {author} {\bibfnamefont {R.~O.}\ \bibnamefont
  {Umucalilar}}\ and\ \bibinfo {author} {\bibfnamefont {I.}~\bibnamefont
  {Carusotto}},\ }\bibfield  {title} {\enquote {\bibinfo {title} {Artificial
  gauge field for photons in coupled cavity arrays},}\ }\href {\doibase
  10.1103/PhysRevA.84.043804} {\bibfield  {journal} {\bibinfo  {journal} {Phys.
  Rev. A}\ }\textbf {\bibinfo {volume} {84}},\ \bibinfo {pages} {043804--8}
  (\bibinfo {year} {2011})}\BibitemShut {NoStop}%
\bibitem [{\citenamefont {Fang}\ \emph {et~al.}(2012)\citenamefont {Fang},
  \citenamefont {Yu},\ and\ \citenamefont {Fan}}]{Fang:2012-782:NPHOT}%
  \BibitemOpen
  \bibfield  {author} {\bibinfo {author} {\bibfnamefont {K.~J.}\ \bibnamefont
  {Fang}}, \bibinfo {author} {\bibfnamefont {Z.~F.}\ \bibnamefont {Yu}}, \ and\
  \bibinfo {author} {\bibfnamefont {S.~H.}\ \bibnamefont {Fan}},\ }\bibfield
  {title} {\enquote {\bibinfo {title} {Realizing effective magnetic field for
  photons by controlling the phase of dynamic modulation},}\ }\href {\doibase
  10.1038/NPHOTON.2012.236} {\bibfield  {journal} {\bibinfo  {journal} {Nature
  Photonics}\ }\textbf {\bibinfo {volume} {6}},\ \bibinfo {pages} {782--787}
  (\bibinfo {year} {2012})}\BibitemShut {NoStop}%
\bibitem [{\citenamefont {Khanikaev}\ \emph {et~al.}(2013)\citenamefont
  {Khanikaev}, \citenamefont {Mousavi}, \citenamefont {Tse}, \citenamefont
  {Kargarian}, \citenamefont {MacDonald},\ and\ \citenamefont
  {Shvets}}]{Khanikaev:2013-233:NMAT}%
  \BibitemOpen
  \bibfield  {author} {\bibinfo {author} {\bibfnamefont {A.~B.}\ \bibnamefont
  {Khanikaev}}, \bibinfo {author} {\bibfnamefont {S.~H.}\ \bibnamefont
  {Mousavi}}, \bibinfo {author} {\bibfnamefont {W.~K.}\ \bibnamefont {Tse}},
  \bibinfo {author} {\bibfnamefont {M.}~\bibnamefont {Kargarian}}, \bibinfo
  {author} {\bibfnamefont {A.~H.}\ \bibnamefont {MacDonald}}, \ and\ \bibinfo
  {author} {\bibfnamefont {G.}~\bibnamefont {Shvets}},\ }\bibfield  {title}
  {\enquote {\bibinfo {title} {Photonic topological insulators},}\ }\href
  {\doibase 10.1038/NMAT3520} {\bibfield  {journal} {\bibinfo  {journal} {Nat.
  Mater.}\ }\textbf {\bibinfo {volume} {12}},\ \bibinfo {pages} {233--239}
  (\bibinfo {year} {2013})}\BibitemShut {NoStop}%
\bibitem [{\citenamefont {Kraus}\ \emph {et~al.}(2012)\citenamefont {Kraus},
  \citenamefont {Lahini}, \citenamefont {Ringel}, \citenamefont {Verbin},\ and\
  \citenamefont {Zilberberg}}]{Kraus:2012-106402:PRL}%
  \BibitemOpen
  \bibfield  {author} {\bibinfo {author} {\bibfnamefont {Y.~E.}\ \bibnamefont
  {Kraus}}, \bibinfo {author} {\bibfnamefont {Y.}~\bibnamefont {Lahini}},
  \bibinfo {author} {\bibfnamefont {Z.}~\bibnamefont {Ringel}}, \bibinfo
  {author} {\bibfnamefont {M.}~\bibnamefont {Verbin}}, \ and\ \bibinfo {author}
  {\bibfnamefont {O.}~\bibnamefont {Zilberberg}},\ }\bibfield  {title}
  {\enquote {\bibinfo {title} {Topological states and adiabatic pumping in
  quasicrystals},}\ }\href {\doibase 10.1103/PhysRevLett.109.106402} {\bibfield
   {journal} {\bibinfo  {journal} {Phys. Rev. Lett.}\ }\textbf {\bibinfo
  {volume} {109}},\ \bibinfo {pages} {106402--5} (\bibinfo {year}
  {2012})}\BibitemShut {NoStop}%
\bibitem [{\citenamefont {Rechtsman}\ \emph {et~al.}(2013)\citenamefont
  {Rechtsman}, \citenamefont {Zeuner}, \citenamefont {Plotnik}, \citenamefont
  {Lumer}, \citenamefont {Podolsky}, \citenamefont {Dreisow}, \citenamefont
  {Nolte}, \citenamefont {Segev},\ and\ \citenamefont
  {Szameit}}]{Rechtsman:2013-196:NAT}%
  \BibitemOpen
  \bibfield  {author} {\bibinfo {author} {\bibfnamefont {M.~C.}\ \bibnamefont
  {Rechtsman}}, \bibinfo {author} {\bibfnamefont {J.~M.}\ \bibnamefont
  {Zeuner}}, \bibinfo {author} {\bibfnamefont {Y.}~\bibnamefont {Plotnik}},
  \bibinfo {author} {\bibfnamefont {Y.}~\bibnamefont {Lumer}}, \bibinfo
  {author} {\bibfnamefont {D.}~\bibnamefont {Podolsky}}, \bibinfo {author}
  {\bibfnamefont {F.}~\bibnamefont {Dreisow}}, \bibinfo {author} {\bibfnamefont
  {S.}~\bibnamefont {Nolte}}, \bibinfo {author} {\bibfnamefont
  {M.}~\bibnamefont {Segev}}, \ and\ \bibinfo {author} {\bibfnamefont
  {A.}~\bibnamefont {Szameit}},\ }\bibfield  {title} {\enquote {\bibinfo
  {title} {Photonic {F}loquet topological insulators},}\ }\href {\doibase
  10.1038/nature12066} {\bibfield  {journal} {\bibinfo  {journal} {Nature}\
  }\textbf {\bibinfo {volume} {496}},\ \bibinfo {pages} {196--200} (\bibinfo
  {year} {2013})}\BibitemShut {NoStop}%
\bibitem [{\citenamefont {Hafezi}\ \emph {et~al.}(2013)\citenamefont {Hafezi},
  \citenamefont {Mittal}, \citenamefont {Fan}, \citenamefont {Migdall},\ and\
  \citenamefont {Taylor}}]{Hafezi:2013-1001:NPHOT}%
  \BibitemOpen
  \bibfield  {author} {\bibinfo {author} {\bibfnamefont {M.}~\bibnamefont
  {Hafezi}}, \bibinfo {author} {\bibfnamefont {S.}~\bibnamefont {Mittal}},
  \bibinfo {author} {\bibfnamefont {J.}~\bibnamefont {Fan}}, \bibinfo {author}
  {\bibfnamefont {A.}~\bibnamefont {Migdall}}, \ and\ \bibinfo {author}
  {\bibfnamefont {J.~M.}\ \bibnamefont {Taylor}},\ }\bibfield  {title}
  {\enquote {\bibinfo {title} {Imaging topological edge states in silicon
  photonics},}\ }\href {\doibase 10.1038/NPHOTON.2013.274} {\bibfield
  {journal} {\bibinfo  {journal} {Nature Photonics}\ }\textbf {\bibinfo
  {volume} {7}},\ \bibinfo {pages} {1001--1005} (\bibinfo {year}
  {2013})}\BibitemShut {NoStop}%
\bibitem [{\citenamefont {Liang}\ and\ \citenamefont
  {Chong}(2013)}]{Liang:2013-203904:PRL}%
  \BibitemOpen
  \bibfield  {author} {\bibinfo {author} {\bibfnamefont {G.~Q.}\ \bibnamefont
  {Liang}}\ and\ \bibinfo {author} {\bibfnamefont {Y.~D.}\ \bibnamefont
  {Chong}},\ }\bibfield  {title} {\enquote {\bibinfo {title} {Optical resonator
  analog of a two-dimensional topological insulator},}\ }\href {\doibase
  10.1103/PhysRevLett.110.203904} {\bibfield  {journal} {\bibinfo  {journal}
  {Phys. Rev. Lett.}\ }\textbf {\bibinfo {volume} {110}},\ \bibinfo {pages}
  {203904--5} (\bibinfo {year} {2013})}\BibitemShut {NoStop}%
\bibitem [{\citenamefont {Pasek}\ and\ \citenamefont
  {Chong}(2014)}]{Pasek:2014-75113:PRB}%
  \BibitemOpen
  \bibfield  {author} {\bibinfo {author} {\bibfnamefont {M.}~\bibnamefont
  {Pasek}}\ and\ \bibinfo {author} {\bibfnamefont {Y.~D.}\ \bibnamefont
  {Chong}},\ }\bibfield  {title} {\enquote {\bibinfo {title} {Network models of
  photonic {F}loquet topological insulators},}\ }\href {\doibase
  10.1103/PhysRevB.89.075113} {\bibfield  {journal} {\bibinfo  {journal} {Phys.
  Rev. B}\ }\textbf {\bibinfo {volume} {89}},\ \bibinfo {pages} {075113--10}
  (\bibinfo {year} {2014})}\BibitemShut {NoStop}%
\bibitem [{\citenamefont {Hu}\ \emph {et~al.}(2015)\citenamefont {Hu},
  \citenamefont {Pillay}, \citenamefont {Wu}, \citenamefont {Pasek},
  \citenamefont {Shum},\ and\ \citenamefont {Chong}}]{Hu:2015-11012:PRX}%
  \BibitemOpen
  \bibfield  {author} {\bibinfo {author} {\bibfnamefont {W.~C.}\ \bibnamefont
  {Hu}}, \bibinfo {author} {\bibfnamefont {J.~C.}\ \bibnamefont {Pillay}},
  \bibinfo {author} {\bibfnamefont {K.}~\bibnamefont {Wu}}, \bibinfo {author}
  {\bibfnamefont {M.}~\bibnamefont {Pasek}}, \bibinfo {author} {\bibfnamefont
  {P.~P.}\ \bibnamefont {Shum}}, \ and\ \bibinfo {author} {\bibfnamefont
  {Y.~D.}\ \bibnamefont {Chong}},\ }\bibfield  {title} {\enquote {\bibinfo
  {title} {Measurement of a topological edge invariant in a microwave
  network},}\ }\href {\doibase 10.1103/PhysRevX.5.011012} {\bibfield  {journal}
  {\bibinfo  {journal} {Phys. Rev. X}\ }\textbf {\bibinfo {volume} {5}},\
  \bibinfo {pages} {011012--7} (\bibinfo {year} {2015})}\BibitemShut {NoStop}%
\bibitem [{\citenamefont {Wu}\ and\ \citenamefont
  {Hu}(2015)}]{Wu:2015-223901:PRL}%
  \BibitemOpen
  \bibfield  {author} {\bibinfo {author} {\bibfnamefont {L.~H.}\ \bibnamefont
  {Wu}}\ and\ \bibinfo {author} {\bibfnamefont {X.}~\bibnamefont {Hu}},\
  }\bibfield  {title} {\enquote {\bibinfo {title} {Scheme for achieving a
  topological photonic crystal by using dielectric material},}\ }\href
  {\doibase 10.1103/PhysRevLett.114.223901} {\bibfield  {journal} {\bibinfo
  {journal} {Phys. Rev. Lett.}\ }\textbf {\bibinfo {volume} {114}},\ \bibinfo
  {pages} {223901--5} (\bibinfo {year} {2015})}\BibitemShut {NoStop}%
\bibitem [{\citenamefont {He}\ \emph {et~al.}(2016)\citenamefont {He},
  \citenamefont {Sun}, \citenamefont {Liu}, \citenamefont {Lu}, \citenamefont
  {Chen}, \citenamefont {Feng},\ and\ \citenamefont
  {Chen}}]{He:2016-4924:PNAS}%
  \BibitemOpen
  \bibfield  {author} {\bibinfo {author} {\bibfnamefont {C.}~\bibnamefont
  {He}}, \bibinfo {author} {\bibfnamefont {X.~C.}\ \bibnamefont {Sun}},
  \bibinfo {author} {\bibfnamefont {X.~P.}\ \bibnamefont {Liu}}, \bibinfo
  {author} {\bibfnamefont {M.~H.}\ \bibnamefont {Lu}}, \bibinfo {author}
  {\bibfnamefont {Y.}~\bibnamefont {Chen}}, \bibinfo {author} {\bibfnamefont
  {L.}~\bibnamefont {Feng}}, \ and\ \bibinfo {author} {\bibfnamefont {Y.~F.}\
  \bibnamefont {Chen}},\ }\bibfield  {title} {\enquote {\bibinfo {title}
  {Photonic topological insulator with broken time-reversal symmetry},}\ }\href
  {\doibase 10.1073/pnas.1525502113} {\bibfield  {journal} {\bibinfo  {journal}
  {Proc. Natl. Acad. Sci. U. S. A.}\ }\textbf {\bibinfo {volume} {113}},\
  \bibinfo {pages} {4924--4928} (\bibinfo {year} {2016})}\BibitemShut {NoStop}%
\bibitem [{\citenamefont {Gao}\ \emph {et~al.}(2016)\citenamefont {Gao},
  \citenamefont {Gao}, \citenamefont {Shi}, \citenamefont {Yang}, \citenamefont
  {Lin}, \citenamefont {Xu}, \citenamefont {Joannopoulos}, \citenamefont
  {Soljacic}, \citenamefont {Chen}, \citenamefont {Lu}, \citenamefont {Chong},\
  and\ \citenamefont {Zhang}}]{Gao:2016-11619:NCOM}%
  \BibitemOpen
  \bibfield  {author} {\bibinfo {author} {\bibfnamefont {F.}~\bibnamefont
  {Gao}}, \bibinfo {author} {\bibfnamefont {Z.}~\bibnamefont {Gao}}, \bibinfo
  {author} {\bibfnamefont {X.~H.}\ \bibnamefont {Shi}}, \bibinfo {author}
  {\bibfnamefont {Z.~J.}\ \bibnamefont {Yang}}, \bibinfo {author}
  {\bibfnamefont {X.}~\bibnamefont {Lin}}, \bibinfo {author} {\bibfnamefont
  {H.~Y.}\ \bibnamefont {Xu}}, \bibinfo {author} {\bibfnamefont {J.~D.}\
  \bibnamefont {Joannopoulos}}, \bibinfo {author} {\bibfnamefont
  {M.}~\bibnamefont {Soljacic}}, \bibinfo {author} {\bibfnamefont {H.~S.}\
  \bibnamefont {Chen}}, \bibinfo {author} {\bibfnamefont {L.}~\bibnamefont
  {Lu}}, \bibinfo {author} {\bibfnamefont {Y.~D.}\ \bibnamefont {Chong}}, \
  and\ \bibinfo {author} {\bibfnamefont {B.~L.}\ \bibnamefont {Zhang}},\
  }\bibfield  {title} {\enquote {\bibinfo {title} {Probing topological
  protection using a designer surface plasmon structure},}\ }\href {\doibase
  10.1038/ncomms11619} {\bibfield  {journal} {\bibinfo  {journal} {Nat.
  Commun.}\ }\textbf {\bibinfo {volume} {7}},\ \bibinfo {pages} {11619--9}
  (\bibinfo {year} {2016})}\BibitemShut {NoStop}%
\bibitem [{\citenamefont {Lumer}\ \emph {et~al.}(2013)\citenamefont {Lumer},
  \citenamefont {Plotnik}, \citenamefont {Rechtsman},\ and\ \citenamefont
  {Segev}}]{Lumer:2013-243905:PRL}%
  \BibitemOpen
  \bibfield  {author} {\bibinfo {author} {\bibfnamefont {Y.}~\bibnamefont
  {Lumer}}, \bibinfo {author} {\bibfnamefont {Y.}~\bibnamefont {Plotnik}},
  \bibinfo {author} {\bibfnamefont {M.~C.}\ \bibnamefont {Rechtsman}}, \ and\
  \bibinfo {author} {\bibfnamefont {M.}~\bibnamefont {Segev}},\ }\bibfield
  {title} {\enquote {\bibinfo {title} {Self-localized states in photonic
  topological insulators},}\ }\href {\doibase 10.1103/PhysRevLett.111.243905}
  {\bibfield  {journal} {\bibinfo  {journal} {Phys. Rev. Lett.}\ }\textbf
  {\bibinfo {volume} {111}},\ \bibinfo {pages} {243905--5} (\bibinfo {year}
  {2013})}\BibitemShut {NoStop}%
\bibitem [{\citenamefont {Zeuner}\ \emph {et~al.}(2015)\citenamefont {Zeuner},
  \citenamefont {Rechtsman}, \citenamefont {Plotnik}, \citenamefont {Lumer},
  \citenamefont {Nolte}, \citenamefont {Rudner}, \citenamefont {Segev},\ and\
  \citenamefont {Szameit}}]{Zeuner:2015-40402:PRL}%
  \BibitemOpen
  \bibfield  {author} {\bibinfo {author} {\bibfnamefont {J.~M.}\ \bibnamefont
  {Zeuner}}, \bibinfo {author} {\bibfnamefont {M.~C.}\ \bibnamefont
  {Rechtsman}}, \bibinfo {author} {\bibfnamefont {Y.}~\bibnamefont {Plotnik}},
  \bibinfo {author} {\bibfnamefont {Y.}~\bibnamefont {Lumer}}, \bibinfo
  {author} {\bibfnamefont {S.}~\bibnamefont {Nolte}}, \bibinfo {author}
  {\bibfnamefont {M.~S.}\ \bibnamefont {Rudner}}, \bibinfo {author}
  {\bibfnamefont {M.}~\bibnamefont {Segev}}, \ and\ \bibinfo {author}
  {\bibfnamefont {A.}~\bibnamefont {Szameit}},\ }\bibfield  {title} {\enquote
  {\bibinfo {title} {Observation of a topological transition in the bulk of a
  non-{H}ermitian system},}\ }\href {\doibase 10.1103/PhysRevLett.115.040402}
  {\bibfield  {journal} {\bibinfo  {journal} {Phys. Rev. Lett.}\ }\textbf
  {\bibinfo {volume} {115}},\ \bibinfo {pages} {040402--5} (\bibinfo {year}
  {2015})}\BibitemShut {NoStop}%
\bibitem [{\citenamefont {Rudner}\ \emph {et~al.}(2013)\citenamefont {Rudner},
  \citenamefont {Lindner}, \citenamefont {Berg},\ and\ \citenamefont
  {Levin}}]{Rudner:2013-31005:PRX}%
  \BibitemOpen
  \bibfield  {author} {\bibinfo {author} {\bibfnamefont {M.~S.}\ \bibnamefont
  {Rudner}}, \bibinfo {author} {\bibfnamefont {N.~H.}\ \bibnamefont {Lindner}},
  \bibinfo {author} {\bibfnamefont {E.}~\bibnamefont {Berg}}, \ and\ \bibinfo
  {author} {\bibfnamefont {M.}~\bibnamefont {Levin}},\ }\bibfield  {title}
  {\enquote {\bibinfo {title} {Anomalous edge states and the bulk-edge
  correspondence for periodically driven two-dimensional systems},}\ }\href
  {\doibase 10.1103/PhysRevX.3.031005} {\bibfield  {journal} {\bibinfo
  {journal} {Phys. Rev. X}\ }\textbf {\bibinfo {volume} {3}},\ \bibinfo {pages}
  {031005--15} (\bibinfo {year} {2013})}\BibitemShut {NoStop}%
\bibitem [{\citenamefont {Gomez-Leon}\ and\ \citenamefont
  {Platero}(2013)}]{Gomez-Leon:2013-200403:PRL}%
  \BibitemOpen
  \bibfield  {author} {\bibinfo {author} {\bibfnamefont {A.}~\bibnamefont
  {Gomez-Leon}}\ and\ \bibinfo {author} {\bibfnamefont {G.}~\bibnamefont
  {Platero}},\ }\bibfield  {title} {\enquote {\bibinfo {title}
  {{F}loquet-{B}loch theory and topology in periodically driven lattices},}\
  }\href {\doibase 10.1103/PhysRevLett.110.200403} {\bibfield  {journal}
  {\bibinfo  {journal} {Phys. Rev. Lett.}\ }\textbf {\bibinfo {volume} {110}},\
  \bibinfo {pages} {200403--5} (\bibinfo {year} {2013})}\BibitemShut {NoStop}%
\bibitem [{\citenamefont {Slobozhanyuk}\ \emph {et~al.}(2015)\citenamefont
  {Slobozhanyuk}, \citenamefont {Poddubny}, \citenamefont {Miroshnichenko},
  \citenamefont {Belov},\ and\ \citenamefont
  {Kivshar}}]{Slobozhanyuk:2015-123901:PRL}%
  \BibitemOpen
  \bibfield  {author} {\bibinfo {author} {\bibfnamefont {A.~P.}\ \bibnamefont
  {Slobozhanyuk}}, \bibinfo {author} {\bibfnamefont {A.~N.}\ \bibnamefont
  {Poddubny}}, \bibinfo {author} {\bibfnamefont {A.~E.}\ \bibnamefont
  {Miroshnichenko}}, \bibinfo {author} {\bibfnamefont {P.~A.}\ \bibnamefont
  {Belov}}, \ and\ \bibinfo {author} {\bibfnamefont {{Yu}.~S.}\ \bibnamefont
  {Kivshar}},\ }\bibfield  {title} {\enquote {\bibinfo {title} {Subwavelength
  topological edge states in optically resonant dielectric structures},}\
  }\href {\doibase 10.1103/PhysRevLett.114.123901} {\bibfield  {journal}
  {\bibinfo  {journal} {Phys. Rev. Lett.}\ }\textbf {\bibinfo {volume} {114}},\
  \bibinfo {pages} {123901--5} (\bibinfo {year} {2015})}\BibitemShut {NoStop}%
\bibitem [{\citenamefont {Poli}\ \emph {et~al.}(2015)\citenamefont {Poli},
  \citenamefont {Bellec}, \citenamefont {Kuhl}, \citenamefont {Mortessagne},\
  and\ \citenamefont {Schomerus}}]{Poli:2015-6710:NCOM}%
  \BibitemOpen
  \bibfield  {author} {\bibinfo {author} {\bibfnamefont {C.}~\bibnamefont
  {Poli}}, \bibinfo {author} {\bibfnamefont {M.}~\bibnamefont {Bellec}},
  \bibinfo {author} {\bibfnamefont {U.}~\bibnamefont {Kuhl}}, \bibinfo {author}
  {\bibfnamefont {F.}~\bibnamefont {Mortessagne}}, \ and\ \bibinfo {author}
  {\bibfnamefont {H.}~\bibnamefont {Schomerus}},\ }\bibfield  {title} {\enquote
  {\bibinfo {title} {Selective enhancement of topologically induced interface
  states in a dielectric resonator chain},}\ }\href {\doibase
  10.1038/ncomms7710} {\bibfield  {journal} {\bibinfo  {journal} {Nat.
  Commun.}\ }\textbf {\bibinfo {volume} {6}},\ \bibinfo {pages} {6710--5}
  (\bibinfo {year} {2015})}\BibitemShut {NoStop}%
\bibitem [{\citenamefont {Leykam}\ \emph {et~al.}(2016)\citenamefont {Leykam},
  \citenamefont {Rechtsman},\ and\ \citenamefont
  {Chong}}]{Leykam:2016-13902:PRL}%
  \BibitemOpen
  \bibfield  {author} {\bibinfo {author} {\bibfnamefont {D.}~\bibnamefont
  {Leykam}}, \bibinfo {author} {\bibfnamefont {M.~C.}\ \bibnamefont
  {Rechtsman}}, \ and\ \bibinfo {author} {\bibfnamefont {Y.~D.}\ \bibnamefont
  {Chong}},\ }\bibfield  {title} {\enquote {\bibinfo {title} {Anomalous
  topological phases and unpaired dirac cones in photonic {F}loquet topological
  insulators},}\ }\href {\doibase 10.1103/PhysRevLett.117.013902} {\bibfield
  {journal} {\bibinfo  {journal} {Phys. Rev. Lett.}\ }\textbf {\bibinfo
  {volume} {117}},\ \bibinfo {pages} {013902--6} (\bibinfo {year}
  {2016})}\BibitemShut {NoStop}%
\bibitem [{\citenamefont {Ke}\ \emph {et~al.}(2016)\citenamefont {Ke},
  \citenamefont {Qin}, \citenamefont {Mei}, \citenamefont {Zhong},
  \citenamefont {Kivshar},\ and\ \citenamefont {Lee}}]{Ke:2016-995:LPR}%
  \BibitemOpen
  \bibfield  {author} {\bibinfo {author} {\bibfnamefont {Y.~G.}\ \bibnamefont
  {Ke}}, \bibinfo {author} {\bibfnamefont {X.~Z.}\ \bibnamefont {Qin}},
  \bibinfo {author} {\bibfnamefont {F.}~\bibnamefont {Mei}}, \bibinfo {author}
  {\bibfnamefont {H.~H.}\ \bibnamefont {Zhong}}, \bibinfo {author}
  {\bibfnamefont {{Yu}.~S.}\ \bibnamefont {Kivshar}}, \ and\ \bibinfo {author}
  {\bibfnamefont {C.}~\bibnamefont {Lee}},\ }\bibfield  {title} {\enquote
  {\bibinfo {title} {Topological phase transitions and thouless pumping of
  light in photonic waveguide arrays},}\ }\href {\doibase
  10.1002/lpor.201600119} {\bibfield  {journal} {\bibinfo  {journal} {Laser
  Photon. Rev.}\ }\textbf {\bibinfo {volume} {10}},\ \bibinfo {pages}
  {995--1001} (\bibinfo {year} {2016})}\BibitemShut {NoStop}%
\bibitem [{\citenamefont {Maczewsky}\ \emph {et~al.}(2017)\citenamefont
  {Maczewsky}, \citenamefont {Zeuner}, \citenamefont {Nolte},\ and\
  \citenamefont {Szameit}}]{Maczewsky:2017-13756:NCOM}%
  \BibitemOpen
  \bibfield  {author} {\bibinfo {author} {\bibfnamefont {L.~J.}\ \bibnamefont
  {Maczewsky}}, \bibinfo {author} {\bibfnamefont {J.~M.}\ \bibnamefont
  {Zeuner}}, \bibinfo {author} {\bibfnamefont {S.}~\bibnamefont {Nolte}}, \
  and\ \bibinfo {author} {\bibfnamefont {A.}~\bibnamefont {Szameit}},\
  }\bibfield  {title} {\enquote {\bibinfo {title} {Observation of photonic
  anomalous {F}loquet topological insulators},}\ }\href {\doibase
  10.1038/ncomms13756} {\bibfield  {journal} {\bibinfo  {journal} {Nat.
  Commun.}\ }\textbf {\bibinfo {volume} {8}},\ \bibinfo {pages} {13756--7}
  (\bibinfo {year} {2017})}\BibitemShut {NoStop}%
\bibitem [{\citenamefont {Mukherjee}\ \emph {et~al.}(2017)\citenamefont
  {Mukherjee}, \citenamefont {Spracklen}, \citenamefont {Valiente},
  \citenamefont {Andersson}, \citenamefont {Ohberg}, \citenamefont {Goldman},\
  and\ \citenamefont {Thomson}}]{Mukherjee:2017-13918:NCOM}%
  \BibitemOpen
  \bibfield  {author} {\bibinfo {author} {\bibfnamefont {S.}~\bibnamefont
  {Mukherjee}}, \bibinfo {author} {\bibfnamefont {A.}~\bibnamefont
  {Spracklen}}, \bibinfo {author} {\bibfnamefont {M.}~\bibnamefont {Valiente}},
  \bibinfo {author} {\bibfnamefont {E.}~\bibnamefont {Andersson}}, \bibinfo
  {author} {\bibfnamefont {P.}~\bibnamefont {Ohberg}}, \bibinfo {author}
  {\bibfnamefont {N.}~\bibnamefont {Goldman}}, \ and\ \bibinfo {author}
  {\bibfnamefont {R.~R.}\ \bibnamefont {Thomson}},\ }\bibfield  {title}
  {\enquote {\bibinfo {title} {Experimental observation of anomalous
  topological edge modes in a slowly driven photonic lattice},}\ }\href
  {\doibase 10.1038/ncomms13918} {\bibfield  {journal} {\bibinfo  {journal}
  {Nat. Commun.}\ }\textbf {\bibinfo {volume} {8}},\ \bibinfo {pages}
  {13918--7} (\bibinfo {year} {2017})}\BibitemShut {NoStop}%
\bibitem [{\citenamefont
  {Hatsugai}(1993{\natexlab{a}})}]{Hatsugai:1993-3697:PRL}%
  \BibitemOpen
  \bibfield  {author} {\bibinfo {author} {\bibfnamefont {Y.}~\bibnamefont
  {Hatsugai}},\ }\bibfield  {title} {\enquote {\bibinfo {title} {Chern number
  and edge states in the integer quantum {H}all-effect},}\ }\href {\doibase
  10.1103/PhysRevLett.71.3697} {\bibfield  {journal} {\bibinfo  {journal}
  {Phys. Rev. Lett.}\ }\textbf {\bibinfo {volume} {71}},\ \bibinfo {pages}
  {3697--3700} (\bibinfo {year} {1993}{\natexlab{a}})}\BibitemShut {NoStop}%
\bibitem [{\citenamefont
  {Hatsugai}(1993{\natexlab{b}})}]{Hatsugai:1993-11851:PRB}%
  \BibitemOpen
  \bibfield  {author} {\bibinfo {author} {\bibfnamefont {Y.}~\bibnamefont
  {Hatsugai}},\ }\bibfield  {title} {\enquote {\bibinfo {title} {Edge states in
  the integer quantum {H}all-effect and the {R}iemann surface of the {B}loch
  function},}\ }\href {\doibase 10.1103/PhysRevB.48.11851} {\bibfield
  {journal} {\bibinfo  {journal} {Phys. Rev. B}\ }\textbf {\bibinfo {volume}
  {48}},\ \bibinfo {pages} {11851--11862} (\bibinfo {year}
  {1993}{\natexlab{b}})}\BibitemShut {NoStop}%
\bibitem [{\citenamefont {Hafezi}(2014)}]{Hafezi:2014-210405:PRL}%
  \BibitemOpen
  \bibfield  {author} {\bibinfo {author} {\bibfnamefont {M.}~\bibnamefont
  {Hafezi}},\ }\bibfield  {title} {\enquote {\bibinfo {title} {Measuring
  topological invariants in photonic systems},}\ }\href {\doibase
  10.1103/PhysRevLett.112.210405} {\bibfield  {journal} {\bibinfo  {journal}
  {Phys. Rev. Lett.}\ }\textbf {\bibinfo {volume} {112}},\ \bibinfo {pages}
  {210405--5} (\bibinfo {year} {2014})}\BibitemShut {NoStop}%
\bibitem [{\citenamefont {Mittal}\ \emph {et~al.}(2016)\citenamefont {Mittal},
  \citenamefont {Ganeshan}, \citenamefont {Fan}, \citenamefont {Vaezi},\ and\
  \citenamefont {Hafezi}}]{Mittal:2016-180:NPHOT}%
  \BibitemOpen
  \bibfield  {author} {\bibinfo {author} {\bibfnamefont {S.}~\bibnamefont
  {Mittal}}, \bibinfo {author} {\bibfnamefont {S.}~\bibnamefont {Ganeshan}},
  \bibinfo {author} {\bibfnamefont {J.~Y.}\ \bibnamefont {Fan}}, \bibinfo
  {author} {\bibfnamefont {A.}~\bibnamefont {Vaezi}}, \ and\ \bibinfo {author}
  {\bibfnamefont {M.}~\bibnamefont {Hafezi}},\ }\bibfield  {title} {\enquote
  {\bibinfo {title} {Measurement of topological invariants in a {2D} photonic
  system},}\ }\href {\doibase 10.1038/NPHOTON.2016.10} {\bibfield  {journal}
  {\bibinfo  {journal} {Nature Photonics}\ }\textbf {\bibinfo {volume} {10}},\
  \bibinfo {pages} {180--184} (\bibinfo {year} {2016})}\BibitemShut {NoStop}%
\bibitem [{\citenamefont {Garanovich}\ \emph {et~al.}(2008)\citenamefont
  {Garanovich}, \citenamefont {Sukhorukov},\ and\ \citenamefont
  {Kivshar}}]{Garanovich:2008-203904:PRL}%
  \BibitemOpen
  \bibfield  {author} {\bibinfo {author} {\bibfnamefont {I.~L.}\ \bibnamefont
  {Garanovich}}, \bibinfo {author} {\bibfnamefont {A.~A.}\ \bibnamefont
  {Sukhorukov}}, \ and\ \bibinfo {author} {\bibfnamefont {{Yu}.~S.}\
  \bibnamefont {Kivshar}},\ }\bibfield  {title} {\enquote {\bibinfo {title}
  {Defect-free surface states in modulated photonic lattices},}\ }\href
  {\doibase 10.1103/PhysRevLett.100.203904} {\bibfield  {journal} {\bibinfo
  {journal} {Phys. Rev. Lett.}\ }\textbf {\bibinfo {volume} {100}},\ \bibinfo
  {pages} {203904--4} (\bibinfo {year} {2008})}\BibitemShut {NoStop}%
\bibitem [{\citenamefont {Szameit}\ \emph {et~al.}(2008)\citenamefont
  {Szameit}, \citenamefont {Garanovich}, \citenamefont {Heinrich},
  \citenamefont {Sukhorukov}, \citenamefont {Dreisow}, \citenamefont {Pertsch},
  \citenamefont {Nolte}, \citenamefont {Tunnermann},\ and\ \citenamefont
  {Kivshar}}]{Szameit:2008-203902:PRL}%
  \BibitemOpen
  \bibfield  {author} {\bibinfo {author} {\bibfnamefont {A.}~\bibnamefont
  {Szameit}}, \bibinfo {author} {\bibfnamefont {I.~L.}\ \bibnamefont
  {Garanovich}}, \bibinfo {author} {\bibfnamefont {M.}~\bibnamefont
  {Heinrich}}, \bibinfo {author} {\bibfnamefont {A.~A.}\ \bibnamefont
  {Sukhorukov}}, \bibinfo {author} {\bibfnamefont {F.}~\bibnamefont {Dreisow}},
  \bibinfo {author} {\bibfnamefont {T.}~\bibnamefont {Pertsch}}, \bibinfo
  {author} {\bibfnamefont {S.}~\bibnamefont {Nolte}}, \bibinfo {author}
  {\bibfnamefont {A.}~\bibnamefont {Tunnermann}}, \ and\ \bibinfo {author}
  {\bibfnamefont {{Yu}.~S.}\ \bibnamefont {Kivshar}},\ }\bibfield  {title}
  {\enquote {\bibinfo {title} {Observation of defect-free surface modes in
  optical waveguide arrays},}\ }\href {\doibase 10.1103/PhysRevLett.101.203902}
  {\bibfield  {journal} {\bibinfo  {journal} {Phys. Rev. Lett.}\ }\textbf
  {\bibinfo {volume} {101}},\ \bibinfo {pages} {203902--4} (\bibinfo {year}
  {2008})}\BibitemShut {NoStop}%
\bibitem [{\citenamefont {Su}\ \emph {et~al.}(1979)\citenamefont {Su},
  \citenamefont {Schrieffer},\ and\ \citenamefont {Heeger}}]{Su:1979-1698:PRL}%
  \BibitemOpen
  \bibfield  {author} {\bibinfo {author} {\bibfnamefont {W.~P.}\ \bibnamefont
  {Su}}, \bibinfo {author} {\bibfnamefont {J.~R.}\ \bibnamefont {Schrieffer}},
  \ and\ \bibinfo {author} {\bibfnamefont {A.~J.}\ \bibnamefont {Heeger}},\
  }\bibfield  {title} {\enquote {\bibinfo {title} {Solitons in
  polyacetylene},}\ }\href {\doibase 10.1103/PhysRevLett.42.1698} {\bibfield
  {journal} {\bibinfo  {journal} {Phys. Rev. Lett.}\ }\textbf {\bibinfo
  {volume} {42}},\ \bibinfo {pages} {1698--1701} (\bibinfo {year}
  {1979})}\BibitemShut {NoStop}%
\bibitem [{\citenamefont {Asb{\'o}th}\ \emph {et~al.}(2016)\citenamefont
  {Asb{\'o}th}, \citenamefont {Oroszl{\'a}ny},\ and\ \citenamefont
  {P{\'a}lyi}}]{asboth2016short}%
  \BibitemOpen
  \bibfield  {author} {\bibinfo {author} {\bibfnamefont {J{\'a}nos~K}\
  \bibnamefont {Asb{\'o}th}}, \bibinfo {author} {\bibfnamefont
  {L{\'a}szl{\'o}}\ \bibnamefont {Oroszl{\'a}ny}}, \ and\ \bibinfo {author}
  {\bibfnamefont {Andr{\'a}s}\ \bibnamefont {P{\'a}lyi}},\ }\bibfield  {title}
  {\enquote {\bibinfo {title} {A short course on topological insulators},}\
  }\href {https://link.springer.com/content/pdf/10.1007/978-3-319-25607-8.pdf}
  {\bibfield  {journal} {\bibinfo  {journal} {Lecture Notes in Physics}\
  }\textbf {\bibinfo {volume} {919}} (\bibinfo {year} {2016})}\BibitemShut
  {NoStop}%
\bibitem [{\citenamefont {Kivshar}\ and\ \citenamefont
  {Turitsyn}(1994)}]{Kivshar:1994-2536:PRE}%
  \BibitemOpen
  \bibfield  {author} {\bibinfo {author} {\bibfnamefont {{Yu}.~S.}\
  \bibnamefont {Kivshar}}\ and\ \bibinfo {author} {\bibfnamefont {S.~K.}\
  \bibnamefont {Turitsyn}},\ }\bibfield  {title} {\enquote {\bibinfo {title}
  {Spatiotemporal pulse collapse on periodic potentials},}\ }\href
  {http://link.aps.org/abstract/PRE/v49/pR2536} {\bibfield  {journal} {\bibinfo
   {journal} {Phys. Rev. E}\ }\textbf {\bibinfo {volume} {49}},\ \bibinfo
  {pages} {R2536--R2539} (\bibinfo {year} {1994})}\BibitemShut {NoStop}%
\bibitem [{\citenamefont {Ryu}\ \emph {et~al.}(2010)\citenamefont {Ryu},
  \citenamefont {Schnyder}, \citenamefont {Furusaki},\ and\ \citenamefont
  {Ludwig}}]{ryu2010topological}%
  \BibitemOpen
  \bibfield  {author} {\bibinfo {author} {\bibfnamefont {S.}~\bibnamefont
  {Ryu}}, \bibinfo {author} {\bibfnamefont {A.~P.}\ \bibnamefont {Schnyder}},
  \bibinfo {author} {\bibfnamefont {A.}~\bibnamefont {Furusaki}}, \ and\
  \bibinfo {author} {\bibfnamefont {A.~W.~W.}\ \bibnamefont {Ludwig}},\
  }\bibfield  {title} {\enquote {\bibinfo {title} {Topological insulators and
  superconductors: tenfold way and dimensional hierarchy},}\ }\href
  {http://iopscience.iop.org/article/10.1088/1367-2630/12/6/065010/meta}
  {\bibfield  {journal} {\bibinfo  {journal} {New Journal of Physics}\ }\textbf
  {\bibinfo {volume} {12}},\ \bibinfo {pages} {065010} (\bibinfo {year}
  {2010})}\BibitemShut {NoStop}%
\bibitem [{\citenamefont {Sriram}\ \emph {et~al.}(2013)\citenamefont {Sriram},
  \citenamefont {Sun},\ and\ \citenamefont {Sarma}}]{ganeshan2013topological}%
  \BibitemOpen
  \bibfield  {author} {\bibinfo {author} {\bibfnamefont {G.}~\bibnamefont
  {Sriram}}, \bibinfo {author} {\bibfnamefont {K.}~\bibnamefont {Sun}}, \ and\
  \bibinfo {author} {\bibfnamefont {S.~D.}\ \bibnamefont {Sarma}},\ }\bibfield
  {title} {\enquote {\bibinfo {title} {Topological zero-energy modes in gapless
  commensurate aubry-andr{\'e}-harper models},}\ }\href {\doibase
  10.1103/PhysRevLett.110.180403} {\bibfield  {journal} {\bibinfo  {journal}
  {Phys. Rev. Lett.}\ }\textbf {\bibinfo {volume} {110}},\ \bibinfo {pages}
  {180403} (\bibinfo {year} {2013})}\BibitemShut {NoStop}%
\bibitem [{\citenamefont {Malkova}\ \emph
  {et~al.}(2009{\natexlab{a}})\citenamefont {Malkova}, \citenamefont {Hromada},
  \citenamefont {Wang}, \citenamefont {Bryant},\ and\ \citenamefont
  {Chen}}]{malkova2009transition}%
  \BibitemOpen
  \bibfield  {author} {\bibinfo {author} {\bibfnamefont {N.}~\bibnamefont
  {Malkova}}, \bibinfo {author} {\bibfnamefont {I.}~\bibnamefont {Hromada}},
  \bibinfo {author} {\bibfnamefont {X.~S.}\ \bibnamefont {Wang}}, \bibinfo
  {author} {\bibfnamefont {G.}~\bibnamefont {Bryant}}, \ and\ \bibinfo {author}
  {\bibfnamefont {Z.~G.}\ \bibnamefont {Chen}},\ }\bibfield  {title} {\enquote
  {\bibinfo {title} {Transition between tamm-like and shockley-like surface
  states in optically induced photonic superlattices},}\ }\href {\doibase
  10.1103/PhysRevA.80.043806} {\bibfield  {journal} {\bibinfo  {journal} {Phys.
  Rev. A}\ }\textbf {\bibinfo {volume} {80}},\ \bibinfo {pages} {043806}
  (\bibinfo {year} {2009}{\natexlab{a}})}\BibitemShut {NoStop}%
\bibitem [{\citenamefont {Malkova}\ and\ \citenamefont
  {Ning}(2007)}]{malkova2007interplay}%
  \BibitemOpen
  \bibfield  {author} {\bibinfo {author} {\bibfnamefont {N.}~\bibnamefont
  {Malkova}}\ and\ \bibinfo {author} {\bibfnamefont {C.~Z.}\ \bibnamefont
  {Ning}},\ }\bibfield  {title} {\enquote {\bibinfo {title} {Interplay between
  tamm-like and shockley-like surface states in photonic crystals},}\ }\href
  {\doibase 10.1103/PhysRevB.76.045305} {\bibfield  {journal} {\bibinfo
  {journal} {Phys. Rev. B}\ }\textbf {\bibinfo {volume} {76}},\ \bibinfo
  {pages} {045305} (\bibinfo {year} {2007})}\BibitemShut {NoStop}%
\bibitem [{\citenamefont {Malkova}\ \emph
  {et~al.}(2009{\natexlab{b}})\citenamefont {Malkova}, \citenamefont {Hromada},
  \citenamefont {Wang}, \citenamefont {Bryant},\ and\ \citenamefont
  {Chen}}]{malkova2009observation}%
  \BibitemOpen
  \bibfield  {author} {\bibinfo {author} {\bibfnamefont {N.}~\bibnamefont
  {Malkova}}, \bibinfo {author} {\bibfnamefont {I.}~\bibnamefont {Hromada}},
  \bibinfo {author} {\bibfnamefont {X.~S.}\ \bibnamefont {Wang}}, \bibinfo
  {author} {\bibfnamefont {G.}~\bibnamefont {Bryant}}, \ and\ \bibinfo {author}
  {\bibfnamefont {Z.~G.}\ \bibnamefont {Chen}},\ }\bibfield  {title} {\enquote
  {\bibinfo {title} {Observation of optical shockley-like surface states in
  photonic superlattices},}\ }\href {\doibase 10.1364/OL.34.001633} {\bibfield
  {journal} {\bibinfo  {journal} {Optics letters}\ }\textbf {\bibinfo {volume}
  {34}},\ \bibinfo {pages} {1633--1635} (\bibinfo {year}
  {2009}{\natexlab{b}})}\BibitemShut {NoStop}%
\bibitem [{\citenamefont {Zak}(1989)}]{Zak:1989-2747:PRL}%
  \BibitemOpen
  \bibfield  {author} {\bibinfo {author} {\bibfnamefont {J.}~\bibnamefont
  {Zak}},\ }\bibfield  {title} {\enquote {\bibinfo {title} {Berry's phase for
  energy-bands in solids},}\ }\href {\doibase 10.1103/PhysRevLett.62.2747}
  {\bibfield  {journal} {\bibinfo  {journal} {Phys. Rev. Lett.}\ }\textbf
  {\bibinfo {volume} {62}},\ \bibinfo {pages} {2747--2750} (\bibinfo {year}
  {1989})}\BibitemShut {NoStop}%
\bibitem [{\citenamefont {Atala}\ \emph {et~al.}(2013)\citenamefont {Atala},
  \citenamefont {Aidelsburger}, \citenamefont {Barreiro}, \citenamefont
  {Abanin}, \citenamefont {Kitagawa}, \citenamefont {Demler},\ and\
  \citenamefont {Bloch}}]{Atala:2013-795:NPHYS}%
  \BibitemOpen
  \bibfield  {author} {\bibinfo {author} {\bibfnamefont {M.}~\bibnamefont
  {Atala}}, \bibinfo {author} {\bibfnamefont {M.}~\bibnamefont {Aidelsburger}},
  \bibinfo {author} {\bibfnamefont {J.~T.}\ \bibnamefont {Barreiro}}, \bibinfo
  {author} {\bibfnamefont {D.}~\bibnamefont {Abanin}}, \bibinfo {author}
  {\bibfnamefont {T.}~\bibnamefont {Kitagawa}}, \bibinfo {author}
  {\bibfnamefont {E.}~\bibnamefont {Demler}}, \ and\ \bibinfo {author}
  {\bibfnamefont {I.}~\bibnamefont {Bloch}},\ }\bibfield  {title} {\enquote
  {\bibinfo {title} {Direct measurement of the {Z}ak phase in topological
  {B}loch bands},}\ }\href {\doibase 10.1038/NPHYS2790} {\bibfield  {journal}
  {\bibinfo  {journal} {Nature Physics}\ }\textbf {\bibinfo {volume} {9}},\
  \bibinfo {pages} {795--800} (\bibinfo {year} {2013})}\BibitemShut {NoStop}%
\bibitem [{\citenamefont {Meier}\ \emph {et~al.}(2016)\citenamefont {Meier},
  \citenamefont {An},\ and\ \citenamefont {Gadway}}]{Meier:2016-13986:NCOM}%
  \BibitemOpen
  \bibfield  {author} {\bibinfo {author} {\bibfnamefont {E.~J.}\ \bibnamefont
  {Meier}}, \bibinfo {author} {\bibfnamefont {F.~A.}\ \bibnamefont {An}}, \
  and\ \bibinfo {author} {\bibfnamefont {B.}~\bibnamefont {Gadway}},\
  }\bibfield  {title} {\enquote {\bibinfo {title} {Observation of the
  topological soliton state in the {S}u-{S}chrieffer-{H}eeger model},}\ }\href
  {\doibase 10.1038/ncomms13986} {\bibfield  {journal} {\bibinfo  {journal}
  {Nat. Commun.}\ }\textbf {\bibinfo {volume} {7}},\ \bibinfo {pages}
  {13986--6} (\bibinfo {year} {2016})}\BibitemShut {NoStop}%
\bibitem [{\citenamefont {Cardano}\ \emph {et~al.}(2017)\citenamefont
  {Cardano}, \citenamefont {D'Errico}, \citenamefont {Dauphin}, \citenamefont
  {Maffei}, \citenamefont {Piccirillo}, \citenamefont {de~Lisio}, \citenamefont
  {De~Filippis}, \citenamefont {Cataudella}, \citenamefont {Santamato},
  \citenamefont {Marrucci}, \citenamefont {Lewenstein},\ and\ \citenamefont
  {Massignan}}]{Cardano:2017-15516:NCOM}%
  \BibitemOpen
  \bibfield  {author} {\bibinfo {author} {\bibfnamefont {F.}~\bibnamefont
  {Cardano}}, \bibinfo {author} {\bibfnamefont {A.}~\bibnamefont {D'Errico}},
  \bibinfo {author} {\bibfnamefont {A.}~\bibnamefont {Dauphin}}, \bibinfo
  {author} {\bibfnamefont {M.}~\bibnamefont {Maffei}}, \bibinfo {author}
  {\bibfnamefont {B.}~\bibnamefont {Piccirillo}}, \bibinfo {author}
  {\bibfnamefont {C.}~\bibnamefont {de~Lisio}}, \bibinfo {author}
  {\bibfnamefont {G.}~\bibnamefont {De~Filippis}}, \bibinfo {author}
  {\bibfnamefont {V.}~\bibnamefont {Cataudella}}, \bibinfo {author}
  {\bibfnamefont {E.}~\bibnamefont {Santamato}}, \bibinfo {author}
  {\bibfnamefont {L.}~\bibnamefont {Marrucci}}, \bibinfo {author}
  {\bibfnamefont {M.}~\bibnamefont {Lewenstein}}, \ and\ \bibinfo {author}
  {\bibfnamefont {P.}~\bibnamefont {Massignan}},\ }\bibfield  {title} {\enquote
  {\bibinfo {title} {Detection of {Z}ak phases and topological invariants in a
  chiral quantum walk of twisted photons},}\ }\href {\doibase
  10.1038/ncomms15516} {\bibfield  {journal} {\bibinfo  {journal} {Nat.
  Commun.}\ }\textbf {\bibinfo {volume} {8}},\ \bibinfo {pages} {15516--7}
  (\bibinfo {year} {2017})}\BibitemShut {NoStop}%
\bibitem [{\citenamefont {Flurin}\ \emph {et~al.}(2017)\citenamefont {Flurin},
  \citenamefont {Ramasesh}, \citenamefont {Hacohen-Gourgy}, \citenamefont
  {Martin}, \citenamefont {Yao},\ and\ \citenamefont
  {Siddiqi}}]{Flurin:2017-31023:PRX}%
  \BibitemOpen
  \bibfield  {author} {\bibinfo {author} {\bibfnamefont {E.}~\bibnamefont
  {Flurin}}, \bibinfo {author} {\bibfnamefont {V.~V.}\ \bibnamefont
  {Ramasesh}}, \bibinfo {author} {\bibfnamefont {S.}~\bibnamefont
  {Hacohen-Gourgy}}, \bibinfo {author} {\bibfnamefont {L.~S.}\ \bibnamefont
  {Martin}}, \bibinfo {author} {\bibfnamefont {N.~Y.}\ \bibnamefont {Yao}}, \
  and\ \bibinfo {author} {\bibfnamefont {I.}~\bibnamefont {Siddiqi}},\
  }\bibfield  {title} {\enquote {\bibinfo {title} {Observing topological
  invariants using quantum walks in superconducting circuits},}\ }\href
  {\doibase 10.1103/PhysRevX.7.031023} {\bibfield  {journal} {\bibinfo
  {journal} {Phys. Rev. X}\ }\textbf {\bibinfo {volume} {7}},\ \bibinfo {pages}
  {031023--6} (\bibinfo {year} {2017})}\BibitemShut {NoStop}%
\end{thebibliography}%

\end{document}